\documentclass[aps,prl,reprint,superscriptaddress]{revtex4-1}

\usepackage{graphicx}
\usepackage{dcolumn}
\usepackage{bm}
\usepackage[colorlinks,linkcolor=blue,anchorcolor=blue,citecolor=blue,urlcolor=blue]{hyperref}
\usepackage{epstopdf}
\usepackage{textcomp}
\usepackage{gensymb}
\usepackage{booktabs}
\usepackage{orcidlink}
\usepackage{mwe}
\usepackage{color}
\usepackage{url}
\usepackage{siunitx}
\usepackage{subfigure}
\usepackage{longtable}
\usepackage{array}
\usepackage{booktabs}
\usepackage{verbatim}
\usepackage{float}
\usepackage{tikz,xcolor,hyperref}

\emergencystretch=\maxdimen
\usepackage{dcolumn}
\newcolumntype{x}{D{.}{.}{6.6}}
\newcolumntype{y}{D{.}{.}{4.5}}
\newcolumntype{z}{D{.}{.}{5.7}}
\newcolumntype{f}{D{.}{.}{7.9}}
\newcolumntype{e}{D{.}{.}{5.6}}

\begin{document}

\preprint{APS/123-QED}

\title{Charge radii of neutron-rich scandium isotopes \\ and the seniority symmetry in the $0f_{7/2}$ shell}

\author{S.~W.~Bai\orcidlink{0000-0002-6087-9788}}
\affiliation{School of Physics and State Key Laboratory of Nuclear Physics and Technology, Peking University, Beijing 100871, China}

\author{X.~F.~Yang\orcidlink{0000-0002-1633-4000}}
\email{xiaofei.yang@pku.edu.cn}
\affiliation{School of Physics and State Key Laboratory of Nuclear Physics and Technology, Peking University, Beijing 100871, China}

\author{\'{A}.~Koszor\'{u}s\orcidlink{0000-0001-7959-8786}}
\affiliation{KU Leuven, Instituut voor Kern-en Stralingsfysica, B-3001 Leuven, Belgium}

\author{J.~C.~Berengut}
\affiliation{School of Physics, University of New South Wales, Sydney, NSW 2052, Australia}
\affiliation{UNSW Nuclear Innovation Centre, UNSW Sydney, Kensington, NSW 2052, Australia}

\author{J.~Billowes}
\affiliation{Department of Physics and Astronomy, The University of Manchester, Manchester, M13 9PL, United Kingdom}
 
\author{M.~L.~Bissell}
\affiliation{Department of Physics and Astronomy, The University of Manchester, Manchester, M13 9PL, United Kingdom}
 
\author{K.~Blaum\orcidlink{0000-0003-4468-9316}}
\affiliation{Max-Planck-Institut f\"{u}r Kernphysik, D-69117 Heidelberg, Germany}

\author{A.~Borschevsky\orcidlink{0000-0002-6558-1921}}
\affiliation{Faculty of Science and Engineering, Van Swinderen Institute for Particle Physics and Gravity, University of Groningen, 9747 AG Groningen, The Netherlands}
 
\author{\mbox{P.~Campbell}}
\affiliation{Department of Physics and Astronomy, The University of Manchester, Manchester, M13 9PL, United Kingdom}
 
\author{B.~Cheal\orcidlink{0000-0002-1490-6263}}
\affiliation{Oliver Lodge Laboratory, Oxford Street, University of Liverpool, Liverpool, L69 7ZE, United Kingdom}
 
\author{C.~S.~Devlin\orcidlink{0000-0002-8034-0085}}
\affiliation{Oliver Lodge Laboratory, Oxford Street, University of Liverpool, Liverpool, L69 7ZE, United Kingdom}
 
\author{K.~T.~Flanagan}
\affiliation{Department of Physics and Astronomy, The University of Manchester, Manchester, M13 9PL, United Kingdom}
\affiliation{Photon Science Institute, Alan Turing Building, University of Manchester, Manchester M13 9PY, United Kingdom}
 
\author{R.~F.~Garcia Ruiz\orcidlink{0000-0002-2926-5569}}
\affiliation{Massachusetts Institute of Technology, Cambridge, MA, USA}
\affiliation{Experimental Physics Department, CERN, CH-1211 Geneva 23, Switzerland}

\author{H.~Heylen\orcidlink{0009-0001-7278-2115}}
\affiliation{Experimental Physics Department, CERN, CH-1211 Geneva 23, Switzerland} 
 
\author{J.~D.~Holt\orcidlink{0000-0003-4833-7959}}
\affiliation{TRIUMF, 4004 Wesbrook Mall, Vancouver, BC V6T 2A3, Canada}
\affiliation{Department of Physics, McGill University, 3600 Rue University, Montr\'eal, QC H3A 2T8, Canada}

\author{B.~S.~Hu\orcidlink{0000-0001-8071-158X}}
\affiliation{TRIUMF, 4004 Wesbrook Mall, Vancouver, BC V6T 2A3, Canada}
\affiliation{National Center for Computational Sciences and Physics Division, Oak Ridge National Laboratory, Oak Ridge, Tennessee 37831, USA}
\affiliation{Cyclotron Institute and Department of Physics and Astronomy, Texas A\& M University, College Station, Texas 77843, USA}
 
\author{A.~Kanellakopoulos\orcidlink{0000-0002-6096-6304}}
\affiliation{KU Leuven, Instituut voor Kern-en Stralingsfysica, B-3001 Leuven, Belgium}
 
\author{J.~Kr\"{a}mer}
\affiliation{Institut f\"{u}r Kernphysik, Technische Universität Darmstadt, D-64289 Darmstadt, Germany}
 
\author{V.~Lagaki}
\affiliation{Experimental Physics Department, CERN, CH-1211 Geneva 23, Switzerland}

\author{B.~Maa\ss{}\orcidlink{ 0000-0002-6844-5706}}
\affiliation{Institut f\"{u}r Kernphysik, Technische Universität Darmstadt, D-64289 Darmstadt, Germany} 

\author{S.~Malbrunot-Ettenauer}
\affiliation{TRIUMF, Vancouver, British Columbia, V6T 2A3, Canada}
\affiliation{Experimental Physics Department, CERN, CH-1211 Geneva 23, Switzerland}
 
\author{T.~Miyagi} 
\affiliation{Center for Computational Sciences, University of Tsukuba, 1-1-1 Tennodai, Tsukuba 305-8577, Japan}
\affiliation{Institut f\"{u}r Kernphysik, Technische Universität Darmstadt, D-64289 Darmstadt, Germany}
\affiliation{ExtreMe Matter Institute EMMI, GSI Helmholtzzentrum für Schwerionenforschung GmbH, 64291 Darmstadt, Germany}
\affiliation{Max-Planck-Institut f\"{u}r Kernphysik, D-69117 Heidelberg, Germany}

\author{\mbox{K. K\"{o}nig}} 
\affiliation{Institut f\"{u}r Kernphysik, Technische Universität Darmstadt, D-64289 Darmstadt, Germany}

\author{M.~Kortelainen\orcidlink{0000-0001-6244-764X}}
\affiliation{Department of Physics, University of Jyv\"askyl\"a, PB 35(YFL) FIN-40351 Jyv\"askyl\"a, Finland}

\author{W.~Nazarewicz\orcidlink{0000-0002-8084-7425}} 
\affiliation{Facility for Rare Isotope Beams, Michigan State University, East Lansing, Michigan 48824, USA}
\affiliation{Department of Physics and Astronomy, Michigan State University, East Lansing, Michigan 48824, USA}

\author{R.~Neugart}
\affiliation{Max-Planck-Institut f\"{u}r Kernphysik, D-69117 Heidelberg, Germany}
\affiliation{Institut f\"{u}r Kernchemie, Universit\"{a}t Mainz, D-55128 Mainz, Germany}
 
\author{G.~Neyens\orcidlink{0000-0001-8613-1455}}
\affiliation{KU Leuven, Instituut voor Kern-en Stralingsfysica, B-3001 Leuven, Belgium}
\affiliation{Experimental Physics Department, CERN, CH-1211 Geneva 23, Switzerland}

\author{W.~N\"{o}rtersh\"{a}user\orcidlink{0000-0001-7432-3687}} 
\affiliation{Institut f\"{u}r Kernphysik, Technische Universität Darmstadt, D-64289 Darmstadt, Germany}

\author{P.-G.~Reinhard\orcidlink{0000-0002-4505-1552}}
\affiliation{Institut f{\"u}r Theoretische Physik, Universit{\"a}t Erlangen, Erlangen, Germany}

\author{\mbox{M. L. Reitsma\orcidlink{0000-0002-8255-7480}}}
\affiliation{Faculty of Science and Engineering, Van Swinderen Institute for Particle Physics and Gravity, University of Groningen, 9747 AG Groningen, The Netherlands}

\author{L.~V.~Rodr\'{i}guez\orcidlink{0000-0002-0093-2110}}
\affiliation{Max-Planck-Institut f\"{u}r Kernphysik, D-69117 Heidelberg, Germany}
\affiliation{Experimental Physics Department, CERN, CH-1211 Geneva 23, Switzerland} 
\affiliation{Institut de Physique Nucl\'{e}aire, CNRS-IN2P3, Universit\'{e} Paris-Sud, Universit\'{e} Paris-Saclay, 91406 Orsay, France}

\author{F.~Sommer}
\affiliation{Institut f\"{u}r Kernphysik, Technische Universität Darmstadt, D-64289 Darmstadt, Germany}

\author{Z.~Y.~Xu}
\affiliation{KU Leuven, Instituut voor Kern-en Stralingsfysica, B-3001 Leuven, Belgium}

\date{\today}      

\begin{abstract}

Nuclear charge radii of neutron-rich $^{47-49}$Sc isotopes were measured using collinear laser spectroscopy at CERN-ISOLDE. The new data reveal that the charge radii of scandium isotopes exhibit a distinct trend between $N=20$ and $N=28$, with $^{41}$Sc and $^{49}$Sc isotopes having similar values, mirroring the closeness of the charge radii of $^{40}$Ca and $^{48}$Ca.
Theoretical models that successfully interpret the radii of calcium isotopes could not account for the observed behavior in scandium radii, in particular the reduced odd-even staggering. Remarkably, the inclusion of the new $^{49}$Sc radius data has unveiled a similar trend in the charge radii of $N=28$ isotones and $Z=20$ isotopes when adding the neutrons atop the $^{40}$Ca core and the protons atop the $^{48}$Ca core, respectively. We demonstrate that this trend is consistent with the prediction of the seniority model.
\end{abstract}

\maketitle

Nuclear charge radii have provided important information on nuclear structure and nuclear models~\cite{Li-radii2006,K-radii-np2021,79Zn-PRL2016,Hg-radii2018,Ye2025}. In particular, high-precision charge radii data across long isotopic chains are essential for understanding inter-nucleon forces and the nuclear many-body problem~\cite{Ca-radii-np2016,Cu-radii-np2020,Ni1-prl2022}. Local variations of the charge radii seen in exotic nuclei with large neutron-to-proton asymmetry often indicate specific structure effects, such as halo structure~\cite{Li-radii2006}, shell closures~\cite{Ca-radii-np2016,Ni2-prl2022,Sn-radii-prl2019}, nuclear shape evolution and coexistence~\cite{79Zn-PRL2016,Hg-radii2018, Au-radii2023}. 

The nuclear charge radii evolution in the $Z=20$ region has long been of interest due to its complex and unique characteristics. A key feature is the arch-like charge radii trend in calcium isotopes at $20 \le N\le 28$, with almost identical radii for $^{40}$Ca and $^{48}$Ca and pronounced odd-even staggering (OES)~\cite{Ca-radii-1984,Ca-radii-np2016,Ar-radii-npa2008,K-radii-np2021}. Since its observation in 1984~\cite{Ca-radii-1984}, explaining this radii pattern between $^{40}$Ca and $^{48}$Ca has posed a long-standing challenge for nuclear theory, see Refs.~\cite{Ca-radii-np2019,Reinhard2021} for its current interpretation. Another puzzle is the absence of the expected radii kink at the $N=20$ shell closure, observed in argon ($Z=18$), potassium ($Z=19$) and calcium ($Z=20$) chains~\cite{Ca-radii-1984,Ca-radii-np2016,Ca-radii-np2019, K-radii-prc2015,K-radii-np2021}. This kink, however, emerged in scandium~($Z=21$) radii at $N=20$, with one proton above the $Z=20$ shell closure~\cite{Sc-radii-prl2023}, prompting further theoretical developments. Why do scandium radii at $N=20$ differ so from their neighbors? Do these differences present across the scandium isotopic chain from $N=20$ to $N=28$? How do the radii of $N=28$ isotones with protons in $0f_{7/2}$ orbit compared to $Z=20$ isotopes with neutrons in $0f_{7/2}$ orbit, well-isolated from its neighboring orbitals by magic numbers 20 and 28? To answer these questions, new charge radii measurements for scandium isotopes near the $N=28$ shell closure are required.

In this letter, we report the charge radii measurement of neutron-rich scandium isotopes up to $N=28$ using collinear laser spectroscopy (COLLAPS) at ISOLDE-CERN. Combined with previous data around $N=20$~\cite{Avgoulea-jpg2011,42Sc-agi,Sc-radii-prl2023}, our measurements complete the charge radius evolution of scandium isotopes between $N=20$ and $N=28$. The data reveal a distinct pattern with nearly-identical sizes for $^{41}$Sc and $^{49}$Sc. Additionally, new $^{49}$Sc radius provide a key datum for $N=28$ isotones, where protons occupy the $0f_{7/2}$ orbit. Our results show that the radii of $N=28$ isotones exhibit a similar arch-like proton number dependence, mirroring the neutrons-dependent trend in calcium. The experimental data are confronted with nuclear density functional theory (DFT) and valence-space in-medium similarity renormalization group (VS-IMSRG) calculations.

Details of the experiment have been described previously in Ref.~\cite{Sc-plb}. In brief,  radioactive Sc isotopes were produced by impinging 1.4-GeV protons onto a thick Ta-foil target at ISOLDE-CERN and selectively ionized using the resonance ionization laser ion source~\cite{RILIS}. The ions were extracted, accelerated to 40~keV, and mass-separated with a high-resolution mass separator. They were then cooled and bunched in a gas-filled radio-frequency quadrupole Paul trap~\cite{ISCOOL}, providing ion bunches with a $\sim$5~$\mu$s temporal width after 100-ms accumulation. The bunched ions delivered to COLLAPS setup were resonantly excited via $3d4s~^{3\!}D_1\to3d4p~^{3\!}F_2$ transition using a cw frequency-doubled Ti:Sapphire laser at 364.3~nm. Long-term laser-frequency stabilization was achieved with a high-accuracy wavemeter kept at a constant environmental temperature. The wavemeter is calibrated in real time by a diode laser locked to a hyperfine transition of $^{87}$Rb. The ion velocity was tuned to match the Doppler-shifted laser frequency for the studied Sc$^+$ transition. Fluorescence photons emitted from the laser-excited scandium ions were detected by photomultipliers and counted as a function of the tuning velocity to obtain the hyperfine structure~(hfs) spectra. All isotopes were measured alternately with the reference isotope $^{45}$Sc to compensate for remaining velocity or frequency excursions.

\begin{figure}[t!]
\includegraphics[width=0.45\textwidth]{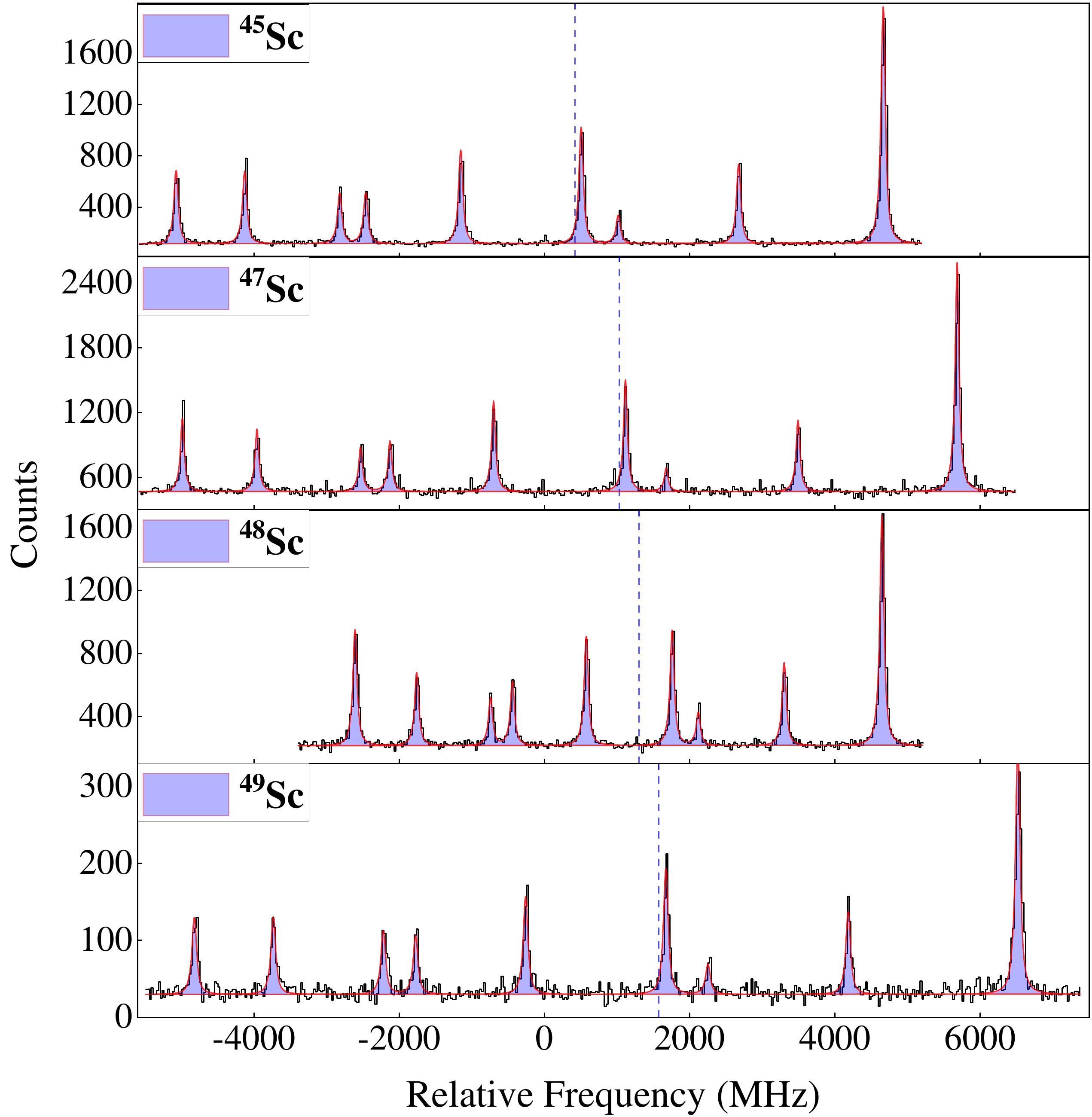}
\vspace{-3mm}
\caption{\footnotesize{Hyperfine structure spectra of $^{45,47-49}$Sc isotopes fitted with a Voigt line profile~(red solid line).}}
\label{fig:fig1}
\vspace{-6mm}
\end{figure}

\begin{table*}[t]
\footnotesize
\caption{\footnotesize{Isotope shifts $\delta \nu^{A,45}$, changes in mean-squared charge radii $\delta \langle r^{2}\rangle^{A,45}$, and charge radii $R_{\rm{ch}}$ of $^{44-49}$Sc obtained in this work~(in bold). Statistical errors are in parentheses, while systematic errors in curly brackets account for the uncertainty in atomic factors $K_{\rm MS}$ and $F$ calculated by atomic theory~\cite{FSCC-2001,CIMBPT}. Value for $\delta \nu^{A,45}$ in the $3d4s$ $^{3\!}D_{2}$ $\to$ $3d4p$ $^{3\!}F_{3}$} transition and $\delta \langle r^{2}\rangle^{A,45}$ for $^{40–46}$Sc are from Refs.~\cite{Avgoulea-jpg2011,42Sc-agi,Sc-radii-prl2023}. For compatibility, the values in column 5 and 6 are recalculated from the reported $\delta \nu^{A,45}$ in literature and atomic $F$ and $K_{\rm MS}$ factors calculated in this work. Note that the spin and parity of the ground and isomeric states of $^{40-49}$Sc are known from NNDC and confirmed through the hfs spectra measurements.}

\renewcommand*{\arraystretch}{1.0}
\begin{ruledtabular}
\begin{tabular}{cc|llll|lll}
& & & \multicolumn{3}{c}{\mbox{$3d4s$ $^{3\!}D_{2}$} $\to$ \mbox{$3d4p$ $^{3\!}F_{3}$}} & \multicolumn{2}{c}{\mbox{$3d4s$ $^{3\!}D_{1}$} $\to$ \mbox{$3d4p$ $^{3\!}F_{2}$}} & \\
\hline
$A$ & $I^{\pi}$& $\delta \nu^{A,45}$& $\delta \langle r^{2}\rangle^{A,45}$~\cite{Sc-radii-prl2023}& \multicolumn{2}{c}{Recalculated}  &$\delta \nu^{A,45}$  & $\delta \langle r^{2}\rangle^{A,45}$ & $R_{\rm ch}$  \\
& &  (MHz) & (fm$^{2}$) &  $\delta \langle r^{2}\rangle^{A,45}$~(fm$^{2}$) &$R_{\rm ch}$~(fm) & (MHz) & (fm$^{2}$) & (fm)  \\
\hline
40 & 4$^{-}$ & $-1594\,(8)$~\cite{Sc-radii-prl2023} & $-$0.226\,(22)\,\{175\} & $-$0.197\,(21)\,\{371\} & 3.518\,(3)\,\{53\}& &  &   \\
41 & 7/2$^{-}$ & $-1199\,(4)$~\cite{Sc-radii-prl2023} & $-$0.305\,(10)\,\{137\} & $-$0.279\,(11)\,\{290\}& 3.506\,(2)\,\{41\} & & &  \\
42 & 0$^{+}$  & $-985\,(11)$~\cite{Avgoulea-jpg2011}  & $+$0.076\,(31)\,\{100\} & $+$0.082\,(29)\,\{212\}& 3.558\,(4)\,\{30\}& & &   \\
42m & 7$^{+}$  & $-911\,(5)$~\cite{42Sc-agi} & $-$0.134\,(34)\,\{100\} & $-$0.115\,(13)\,\{212\} &3.530\,(2)\,\{30\}&  & &   \\
43 & 7/2$^{-}$ & $-631\,(5)$~\cite{Avgoulea-jpg2011} & $+$0.019\,(14)\,\{65\} & $+$0.025\,(13)\,\{138\} & 3.549\,(2)\,\{20\}& & &  \\
44 & 2$^{+}$  & $-287\,(4)$~\cite{Avgoulea-jpg2011} & $-$0.051\,(11)\,\{32\}& $-$0.044\,(11)\,\{68\}& 3.540\,(2)\,\{10\} &\bf{--289.3\,(13)} & \bf{--0.029\,(3)\,\{68\}} & \bf{3.542\,(1)\,\{10\}}  \\
44m & 6$^{+}$  & $-262\,(6)$~\cite{Avgoulea-jpg2011} & $-$0.122\,(16)\,\{32\} & $-$0.111\,(16)\,\{68\} & 3.530\,(2)\,\{10\}  &\bf{--260.0\,(7)} & \bf{--0.107\,(2)\,\{68\}} &  \bf{3.531\,(1)\,\{10\}} \\
45m & 3/2$^{+}$  & $-66\,(2)$~\cite{Avgoulea-jpg2011} & $+$0.187\,(6)\,\{6\}& $+$0.177\,(5)\,\{6\} & 3.571\,(1)\,\{3\}  && &  \\
46 & 4$^{+}$  & $+336\,(3)$~\cite{Avgoulea-jpg2011} & $-$0.124\,(9)\,\{31\} & $-$0.120\,(8)\,\{65\} & 3.529\,(1)\,\{10\} & \bf{+308.8\,(6)} & \bf{--0.056\,(2)\,\{65\}} & \bf{3.538\,(1)\,\{10\}} \\
47 & 7/2$^{-}$ & & & & &\bf{+614.1\,(6)} & \bf{--0.137\,(2)\,\{127\}} & \bf{3.526\,(1)\,\{18\}} \\
48 & 6$^{+}$  & & & & &\bf{+916.2\,(6)} & \bf{--0.239\,(2)\,\{186\}} & \bf{3.512\,(1)\,\{27\}} \\
49 & 7/2$^{-}$ & & & & &\bf{+1186.6\,(11)} & \bf{--0.286\,(3)\,\{243\}} & \bf{3.505\,(1)\,\{35\}} \\
\end{tabular}
\label{tab1}
\end{ruledtabular}
\end{table*}

Figure~\ref{fig:fig1} shows typical hfs spectra for $^{45,47-49}$Sc isotopes, fitted using a Voigt profile in SATLAS package~\cite{SATLAS}. The extracted hfs parameters of odd-$A$ scandium isotopes were reported in Ref.~\cite{Sc-plb}, while those for even-$A$ scandium isotopes will be detailed in a future publication. Isotope shifts $\delta\nu^{A,45}$ = $\nu^{A}$ - $\nu^{45}$ of $^{44-49}$Sc isotopes extracted from center-of-gravity~(COG)~$\nu^{A}$ with respect to $\nu^{45}$ of the reference isotope $^{45}$Sc, are presented in Table~\ref{tab1}. The changes in mean-square charge radii $\delta\langle r^2 \rangle^{A,45}$ could then be deduced via
\begin{equation}
    \delta\nu^{A,45} = K_{\rm MS}\frac{M_{A} - M_{45}}{M_{A}M_{45}} + F\delta\langle r^2\rangle^{A,45}.
\end{equation}
Here, $K_{\rm{MS}}$ and $F$ are the mass-shift and field-shift factors, respectively. And $M_{45}$ and $M_{A}$ are the atomic masses of $^{45}$Sc and $^{A}$Sc, respectively. 

To calculate $\delta\langle r^2 \rangle^{A,45}$, the atomic factors $K_{\rm{MS}}$ and $F$ must be determined from atomic theory as only one stable isotope $^{45}$Sc is available. In this work, $F$ was calculated with the relativistic Fock-space coupled cluster method (FSCC)~\cite{FSCC-1998,FSCC-2001} in the DIRAC program~\cite{DIRAC19}, while $K_{\rm{MS}}$ was predicted using configuration interaction with many-body perturbation theory (CI+MBPT)~\cite{CIMBPT} within the AMBIT program~\cite{AMBIT-2019}. The closed-shell Sc$^{3+}$ [Ar] electron configuration was used for the Dirac-Hartree-Fock reference state for both methods. The two-particle FSCC sector was employed in the $F$ calculations where two electrons were added to the reference state to reach the [Ar]3d4s states of Sc$^{+}$.~The intermediate Hamiltonian (IH) technique~\cite{landau2001intermediate} was used to aid convergence, and the two electrons could occupy orbits in a model space consisting of the $3d 4s 4p (4d 5s 5p 4f 5d 6s 6p)$ orbits, with the IH space in parentheses. Dyall's 4z basis set~\cite{dyallprep3d} consisting of $31s21p13d8f6g3h$ basis functions was used, including core correlating functions and a layer of diffuse functions. All electrons were correlated and the virtual space was cut off at 200~a.u. A Gaussian nuclear charge distribution~\cite{visscher-nuccharge,Sn-Fredik} was used and $F$ was computed from the numerical derivative of energy calculations at different $\left\langle r^2 \right\rangle$ values. For the CI+MBPT calculations, a B-spline basis was used, and CI included single and double excitations from the $3d4s$, $3d^2$, $4s^2$, $3d4p$, $4s4p$, $3d5s$, $3d4d$ configurations up to $20spdfg$, while the MBPT basis consisting of $40spdfgh$ was used to treat core-valence correlations and included up to three-body MBPT diagrams. $K_{\rm{MS}}$ was calculated using the finite field method, where either the normal or specific mass shift operator is added to the Hamiltonian. The uncertainty was estimated by analyzing contributions from different computational parameters~\cite{Sn-Fredik,Ge-moment-prc2020,RaF-prl}, such as the limited size of the basis set, higher-order correlation effects and the relativistic Gaunt term. Further details are in Ref.~\cite{Sn-Fredik} and references therein. 

\begin{table}[!htb]
\caption{\footnotesize{Calculated $F$ and $K_{\rm MS}$ factors for two scandium ion transitions using FSCC and CI+MBPT. For comparison, the atomic factors calculated with MCDHF method~\cite{Sc-radii-prl2023} are also included.}}
\renewcommand*{\arraystretch}{1.0}
\begin{ruledtabular}
\begin{tabular}{l|ll}
 Transition& $F$ (MHz/fm$^{2}$)  & $K_{\rm MS}$ (GHz$\cdot$u) \\
\hline
$3d4s~^{3\!}D_2\to3d4p~^{3\!}F_3$ &  $-373\,(12)$ & $602\,(50)$  \\
 & $-352\,(12)$~\cite{Sc-radii-prl2023} & $604\,(22)$~\cite{Sc-radii-prl2023} \\
 \hline
 $3d4s~^{3\!}D_1\to3d4p~^{3\!}F_2$ & $-373\,(12)$ & $595\,(50)$  \\
\end{tabular}
\end{ruledtabular}
\label{tab2}
\end{table}

\begin{figure}[!htb]
\includegraphics[width=0.43\textwidth]{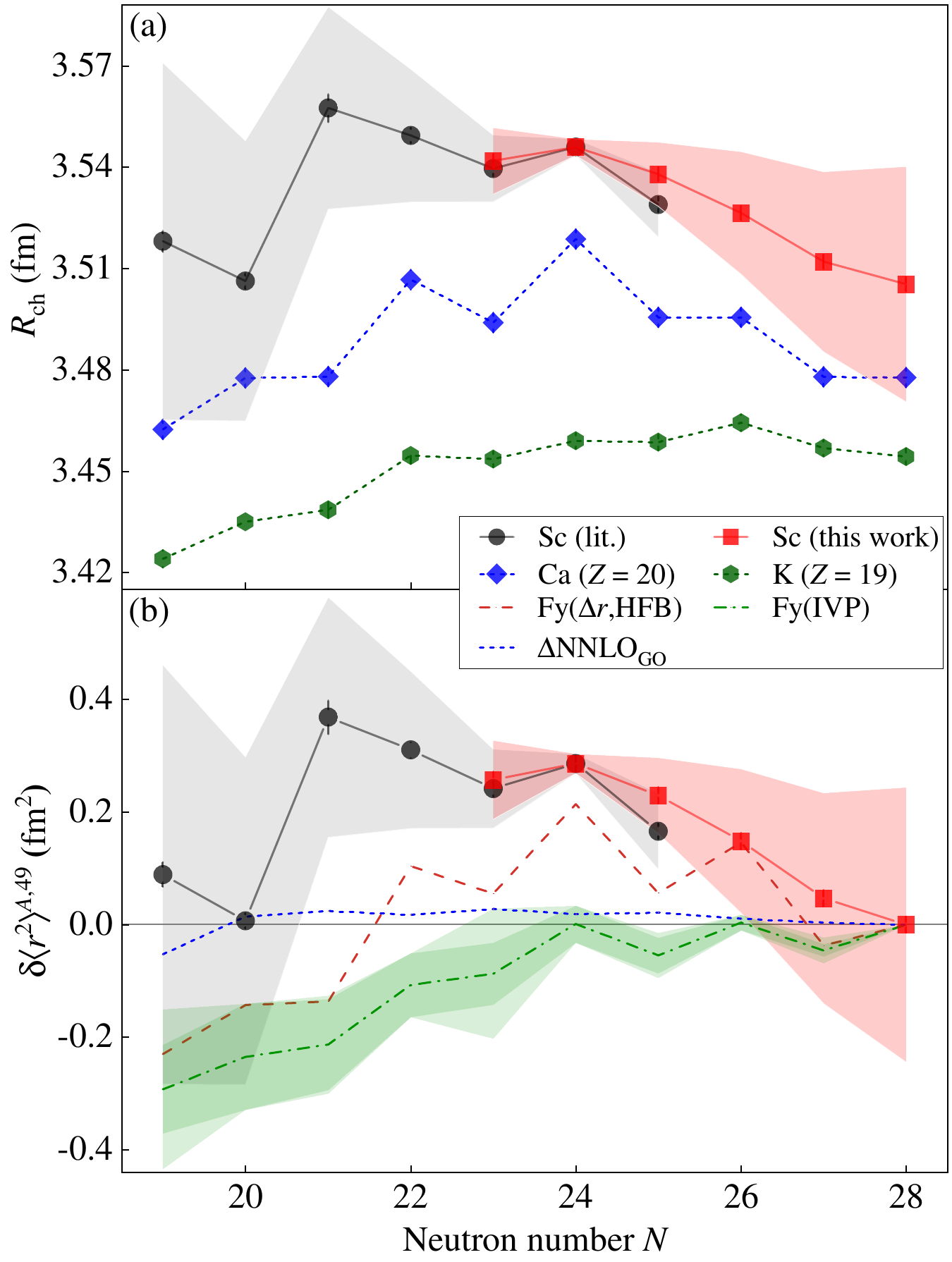}
\vspace{-3mm}
\caption{\footnotesize{(a)~Charge radii $R_{\rm{ch}}$ of $^{40-49}$Sc, $^{39-48}$Ca~\cite{Ca-radii-np2016,Ca-radii-np2019}, and $^{38-47}$K~\cite{K-radii-np2021}. The grey and red shaded bands represent systematic uncertainties in the scandium radii calculated from $\delta\nu^{A,45}$ of $3d4s~^{3\!}D_1$ $\to$ $3d4p~^{3\!}F_2$ and $3d4s~^{3\!}D_2$ $\to$ $3d4p~^{3\!}F_3$ transitions, respectively. These uncertainties arise from the $K_{\rm MS}$ and $F$ factors shown in Tab.~\ref{tab2}, as well as from the uncertainty in reference value $R_{\rm ch}(^{45}\text{Sc}$). For potassium and calcium radii, only statistical errors are included. (b) $\delta \langle r^{2}\rangle^{A,49}$ of $^{40-49}$Sc compared to theoretical predictions from Fy($\Delta r$, HFB), Fy(IVP), and VS-IMSRG using using $\Delta{\rm{NNLO}}_{\rm{GO}}$. The uncertainties for Fy(IVP) results are shown with a green shaded band, with the darker green indicating statistical uncertainties only.}}
\label{fig:fig2}
\vspace{-6mm}
\end{figure}

The $\delta\nu^{A,45}$ of $3d4s~^{3\!}D_2$ $\to$ $3d4p~^{3\!}F_3$ ionic transition are available in the literature for $^{40-46}$Sc isotopes, with the corresponding $K_{\rm MS}$ and $F$ factors calculated using multiconfiguration Dirac-Hartree-Fock (MCDHF) method~\cite{Avgoulea-jpg2011,42Sc-agi,Sc-radii-prl2023}. To obtain a self-consistent set of scandium charge radii combing the $\delta\nu^{A,45}$ measured in this work and those in the literature~\cite{Avgoulea-jpg2011,42Sc-agi,Sc-radii-prl2023}, we performed calculations using FSCC and CI+MBPT for both ionic transitions. The results are shown in Table~\ref{tab2}, together with those used in Ref.~\cite{Sc-radii-prl2023}. Using the newly-calculated $K_{\rm MS}$ and $F$ factors, we obtained $\delta \langle r^{2}\rangle^{A,45}$ and $R_{\rm{ch}}$ for $^{44-49}$Sc isotopes (this work) and reevaluated those for $^{40-46}$Sc~\cite{Avgoulea-jpg2011,42Sc-agi,Sc-radii-prl2023}, as summarized in Table~\ref{tab1}. The $R_{\rm ch}$ of radioactive isotopes was calculated using the reference value $R_{\rm ch}(^{45}\text{Sc}) = 3.5459(25)$~fm~\cite{Radii2013}. The uncertainty in the reference value was taken into account in the systematic error of the derived $R_{\rm ch}$ values~(Table~\ref{tab1}). Charge radii $R_{\rm{ch}}$ and $\delta\langle r^2 \rangle^{A,49}=\delta\langle r^2 \rangle^{A,45}-\delta\langle r^2 \rangle^{49,45}$ of $^{40-49}$Sc are shown in Fig.~\ref{fig:fig2}. A notable discrepancy is observed between the $^{46}$Sc radius calculated from $\delta\nu^{A,45}$~\cite{IGISOL-Sc,Avgoulea-jpg2011} and that obtained in this work. This is likely due to the low statistics of the not-fully-resolved and incomplete hfs spectrum of $^{46}$Sc in Ref.~\cite{IGISOL-Sc}.

With these new results, we can now investigate the overall charge radii behavior of isotopes with \mbox{$Z=20\pm1$} and \mbox{$N=20-28$}, where valence neutrons occupy the $0f_{7/2}$ shell-model orbit. Figure~\ref{fig:fig2}(a) compares the experimental $R_{\rm{ch}}$ of $^{40-49}$Sc, $^{39-48}$Ca~\cite{Ca-radii-np2016,Ca-radii-np2019} and $^{38-47}$K~\cite{K-radii-np2021}. Several new features emerge in addition to the puzzling radii kink at $N=20$ in scandium~\cite{Sc-radii-prl2023} which is not seen in potassium and calcium. Most remarkably, the radii trend of even-$N$ scandium isotopes resembles the one seen in calcium above $N=20$. 
This characteristic trend between $N=20$ and $N=28$ is not observed in the potassium chain. Another notable feature is that the considerable OES seen in calcium charge radii is much less pronounced in both scandium and potassium. 

To get a better understanding on the observed trend in scandium radii, we performed calculations using DFT and VS-IMSRG method.

The DFT calculations were performed using two Fayans energy density functional parametrizations Fy($\Delta r$, HFB)~\cite{Fayan-Ca-prc2017} and Fy(IVP)~\cite{Fy-IVP-2024}. The Fy($\Delta r$, HFB)~\cite{Fayan-Ca-prc2017} has been successful in understanding the charge radii of nearly-closed shell nuclei~(e.g.~K~\cite{K-radii-np2021}, Ca~\cite{Ca-radii-np2019}, Ni~($Z=28$)~\cite{Ni1-prl2022,Ni2-prl2022}, Cu~($Z=29$)~\cite{Cu-radii-np2020}, Cd~($Z=48$)~\cite{Cd-prl2018}, Sn~($Z=50$)~\cite{Sn-radii-prl2019}) and the open-shell nuclei~(e.g.~Ge~($Z=32$)~\cite{Ge-radii-2024} and Pd~($Z=46$)~\cite{Pd-prl2022}). The Fy(IVP) is a new functional aiming to capture the isovector pairing correlations~\cite{Fy-IVP-2024}. The calculations were performed with an axially deformed DFT solver, which allowed nuclear deformations and time-odd polarization for odd-$Z$ and odd-$N$ nuclei. For more details on Fayans DFT methodology and calculations, see Ref.~\cite{Sc-radii-prl2023}. As shown in Fig.~\ref{fig:fig2}(b), the Fy($\Delta r$, HFB) calculations partially capture the $\delta\left\langle r^2 \right\rangle^{N,28}$ trend of scandium when approaching $N=28$, although it overestimates the OES and underestimates the differential radii around $N=20$, as already noted in Ref.~\cite{Sc-radii-prl2023}. The functional Fy(IVP) underestimates both OES and differential radii.

The VS-IMSRG calculations use the $\Delta$NNLO$_{\rm{GO}}$(394)~\cite{NNLOgo} chiral interaction, which has reasonably reproduced the differential charge radii of potassium~\cite{K-radii-np2021} and the semi-magic nickel isotopes~\cite{Ni1-prl2022}. Following the same methodology and parameters as Ref.~\cite{Sc-radii-prl2023}, we decoupled a $1s_{1/2}0d_{3/2}0f_{7/2}1p_{3/2}$ valence-space Hamiltonian above a $^{28}$Si core to capture the influence of cross-shell excitation from $Z=N=20$ magic shell gap. As seen in Fig.~\ref{fig:fig2}(b), VS-IMSRG predicts almost constant radii for scandium between $N=20$ and $N=28$.

Summarizing the results in Fig.~\ref{fig:fig2}(b), we conclude that the 
current nuclear models used to describe neighboring isotopic chains with a considerable success~\cite{K-radii-np2021}, fail in reproducing the observed differential radii trend for scandium isotopes with $19\le N \le 28$. 

\begin{figure}[!htb]
\includegraphics[width=0.5\textwidth]{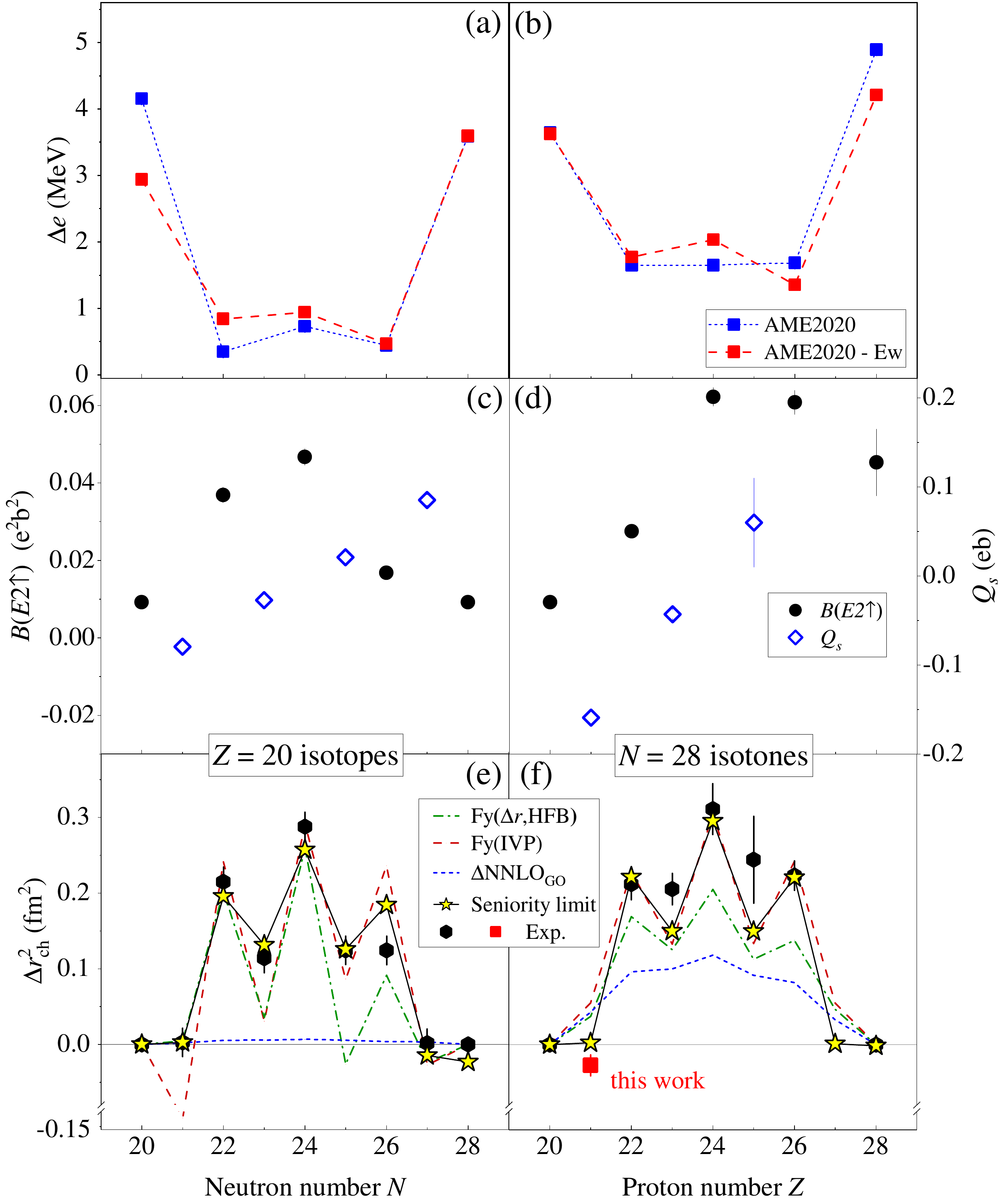}
\vspace{-4mm}
\caption{\footnotesize{(a-b) Single-particle energy-splitting at the Fermi level $\Delta e$~\cite{energy-gap} calculated from experimental one-particle separation energy~\cite{AME2020} with and without the Wigner-energy correction along the $Z=20$ isotopic and $N=28$ isotonic chains. (c-d) Experimental $B(E2\uparrow)$ and $Q_{s}$, taken from NNDC and Refs.~\cite{Sc-plb,Ca-moments2015}, respectively. (e-f) Relative charge radii $\Delta r^{2}_{\rm{ch}}$ for $Z=20$ isotopes and $N=28$ isotones, compared with theoretical calculations. The seniority limit results obtained by fitting the experimental data using Eq.~\ref{rsen} are marked by stars. Experimental radii of $^{40-48}$Ca, $^{50}$Ti, $^{51}$V, $^{52}$Cr, $^{53}$Mn, $^{54}$Fe and $^{56}$Ni are taken from Refs.~\cite{Ca-radii-np2016,Radii2013,Mn-radii,Ni2-prl2022}. See text for details.}}
\label{fig:fig3}
\vspace{-4mm}
\end{figure}

Motivated by the characteristic trend of calcium radii in Fig.~\ref{fig:fig2}(a) with neutrons filling the $0f_{7/2}$ orbit, it is interesting to investigate the charge radii of the semi-magic $N=28$ isotones with protons filling the $0f_{7/2}$ orbit. To gain more insights into the underlying coupling scheme, we show in Fig.~\ref{fig:fig3}(a,b) the single-particle energy splitting at the Fermi level $\Delta e$~\cite{energy-gap,Dobaczewski2001} deduced from experimental one-nucleon separation energies~\cite{AME2020} for $Z=20$ isotopes and $N=28$ isotones. Since $\Delta e$ around $^{40}$Ca and $^{56}$Ni can be affected by the Wigner energy, we show the results with and without the Wigner-energy correction~\cite{energy-gap}. As expected, $\Delta e$ is large at the magic numbers 20 and 28, and is fairly constant and reduced within the $0f_{7/2}$ shell. Figure~\ref{fig:fig3}(c,d) shows the experimental transition probabilities $B(E2\uparrow)$ and spectroscopic quadrupole moments $Q_{s}$. The similarity of the trends reflects the gradual filling of $0f_{7/2}$ neutron and proton orbitals. It is seen that the quadrupole transition probability nicely follow the behavior expected from the seniority scheme. Namely, at mid-shell ($N=24$) the $B(E2\uparrow)$ values reach a maximum and the $Q_{s}$ are close to zero~\cite{Sc-plb,Ca-moments2015,Yang-ppnp2023}.

According to the seniority picture, differential radii for a given semi-magic isotopic chain follow a simple parabolic pattern~\cite{Zamick1971,Talmi1984}:
\begin{equation}\label{rsen}
\delta\langle r_{\rm ch}^2\rangle^{A_{\rm m},A_{\rm m}+n}=
an+bn^2 + c\frac{1}{2}\left[(-1)^n-1 \right],
\end{equation}
where $n$ is the nucleon number above the shell closure at $N=N_{\rm m}$ in the magic nucleus $A_{\rm m}$, and $a$, $b$, and $c$ are constants. The parameter $c$ describes the OES. Recently, the adequacy of the seniority or generalized-seniority expression (\ref{rsen}) was successfully tested in this region~\cite{Ca-Zn-prc2022}. With the $^{49}$Sc radius measured in this work, testing the seniority scheme for charge radii of $Z=20$ isotopes and $N=28$ isotones now becomes possible.

To better visualize the local variations in the radius trend, relative charge radii $\Delta r^{2}_{\rm{ch}}$ was proposed~\cite{beta-corr-prc2022} by subtracting a value obtained from linear interpolation between the two doubly-magic nuclei ($^{40}$Ca, $^{48}$Ca for $Z=20$ isotopes, and $^{48}$Ca, $^{56}$Ni for $N=28$ isotones).
For the $Z=20$ isotopes, $\Delta r^{2}_{\rm{ch}}=r^{2}_A-r^{2}_{^{40}{\rm Ca}}$
as $r_{^{40}{\rm Ca}}\approx r_{^{48}{\rm Ca}}$.
The similarity in charge radii of $^{40}$Ca and $^{48}$Ca comes from the neutron charge distribution and the neutron spin-orbit density, which produce a negative contribution to the charge density in $^{48}$Ca~\cite{Emrich1983,Reinhard2021}.
For the $N=28$ isotones, $r_{^{56}\rm Ni} > r_{^{48}\rm Ca}$ due to the larger mass of $^{56}$Ni and the rapid rise of the spin-orbit charge radius correction between $^{48}$Ca and $^{56}$Ni~\cite{Reinhard2021}. Consequently, we normalize $\Delta r^{2}_{\rm{ch}}$ so that it vanishes for $^{56}$Ni:
\begin{equation}\label{Rnorm}
\Delta r_{\rm{ch}}^2=r^2_{A=48+n}-\left[r^2_{^{48}\rm Ca}+\frac{r^2_{^{56}\rm Ni}-r^2_{^{48}\rm Ca}}{8}\times n\right]. 
\end{equation}
Figures~\ref{fig:fig3}(e,f) show the resulting $\Delta r^{2}_{\rm{ch}}$ trends.
Like in the case of the quadrupole $E2$ operator, the behavior of $\Delta r^{2}_{\rm{ch}}$ is very similar for protons and neutrons occupying the $0f_{7/2}$ shell. 
To further illustrate this point, we optimized the parameters of the seniority-limit expression (\ref{rsen}) to experimental $\Delta r^{2}_{\rm{ch}}$. The resulting parameters $a,b,c$ (all in fm$^2$) are: $a=0.132(7)$, $b=-0.017(1)$, and $c=0.112(10)$ for $Z=20$ isotopes, and $a=0.148(9)$, $b=-0.018(1)$, and $c=0.127(11)$ for $N=28$ isotones. For both chains, the seniority model provides an excellent description of experimental charge radii, consistent with $\Delta e$, $B(E2\uparrow)$ and $Q_{s}$ data. Interestingly, the parameters $a,b,c$ for protons and neutrons in the $0f_{7/2}$ shell agree within the uncertainties, suggesting the proton and neutron monopole polarization effects are very similar.

The behavior of $\Delta r^{2}_{\rm{ch}}$ for $Z=20$ isotopes and $N=28$ isotones is nicely captured by Fy($\Delta r$, HFB) and Fy(IVP) models, consistent with Ref.~\cite{Ca-Zn-prc2022}, but is largely underestimated by the \textit{ab initio} VS-IMSRG calculations with $\Delta{\rm{NNLO}}_{\rm{GO}}$ interaction. This suggests that the valence pairing and quadrupole interactions in our Fayans DFT models align well with the seniority scheme. Despite years of development, current \textit{ab initio} calculations have made only slight improvements in reproducing the distinct radii trends of calcium and scandium isotopes as well as the $N=28$ isotones, as shown in Fig.~\ref{fig:fig2}, Fig.~\ref{fig:fig3} and Refs.~\cite{Sc-radii-prl2023,Ca-Zn-prc2022}. We also note that all the non-implausible chiral interactions from Ref.~\cite{Hu2022} with the same VS-IMSRG calculations failed to reproduce the calcium radii trend. Thus further advancements in many-body methods and additional experimental measurements remain essential.
 
Charge radii of $^{44-49}$Sc isotopes were determined using collinear laser spectroscopy. Together with the data of $^{40-46}$Sc isotopes~\cite{Avgoulea-jpg2011,42Sc-agi,Sc-radii-prl2023}, the scandium radii trend reveals several intricate features: an unexplained behavior around $N=20-22$, nearly equal charge radii for $^{41}$Sc and $^{49}$Sc, and the suppressed OES compared to calcium. Current state-of-the-art nuclear theories, including Fayans DFT and VS-IMSRG, are unable to fully reproduce this trend, highlighting the need for further theoretical improvements. To study the effect of the $0f_{7/2}$ shell occupation in semi-magic $Z=20$ isotopes and $N=28$ isotones, we analyzed the relative charge radii $\Delta r^{2}_{\rm{ch}}$. Remarkably, the radii of $N=28$ isotones exhibit a seniority-limit pattern almost identical to that of the calcium chain. This feature, nicely captured by the DFT calculation using Fy($\Delta r$, HFB) and Fy(IVP), strongly supports the validity of the seniority coupling scheme in this region.

\vspace{1cm}

We acknowledge the support of the ISOLDE collaboration and technical teams. This work was supported in part by the National Key R\&D Program of China (Contract Nos.~2023YFA1606403, 2022YFA1605100), the National Natural Science Foundation of China (Nos: 12027809, 12305122), New Cornerstone Science Foundation through the XPLORER PRIZE; the Max-Planck Society, the BMBF under Contracts No. 05P18RDCIA, and 05P21RDCI1; the BriX Research Program No. P7/12, FWO-Vlaanderen (Belgium) No.~G080022N, GOA 15/010 and C1/22/104 from KU Leuven; the UK Science and Technology Facilities Council grants ST/L005794/1 and ST/P004598/1; ERC Consolidator Grant No.~648381 (FNPMLS); the NSF grant PHY-1068217; the U.S. Department of Energy, Office of Science, Office of Nuclear Physics under grants DE-SC0013365, DE-SC0023175, DE-SC0021176 (Office of Advanced Scientific Computing Research and Office of Nuclear Physics, Scientific Discovery through Advanced Computing), SciDAC-5 (NUCLEI collaboration) and DE-AC05-00OR22725; the Deutsche Forschungsgemeinschaft (DFG, German Research Foundation) – Project-ID 279384907 – SFB 1245, the Helmholtz International Center for FAIR (HIC for FAIR); Dutch Research Council (NWO) project number Vi.Vidi.192.088 and LISA: European Union’s H2020 Framework Programme under grant agreement no. 861198; Academy of Finland under the Academy Project No. 339243; the EU Horizon2020 research and innovation programme through ENSAR2 (No. 654002), NSERC under grants SAPIN-2018-00027 and RGPAS-2018-522453, the Arthur B. McDonald Canadian Astroparticle Physics Research Institute, and JST ERATO Grant No. JPMJER2304, Japan. VS-IMSRG computations were performed with an allocation of computing resources on Cedar at WestGrid and Compute Canada and on Cygnus and Pegasus provided by Multidisciplinary Cooperative Research Program in Center for Computational Sciences, University of Tsukuba. For the FSCC and CI+MBPT calculations we thank the Center for Information Technology of the University of Groningen for their support and for providing access to the Peregrine and H\'abr\'ok high performance computing clusters. we acknowledge the CSC-IT Center for Science Ltd., Finland, for the allocation of computational resources for DFT calculations.

\bibliographystyle{apsrev4-1}
\bibliography{Sc-radii}

\providecommand{\noopsort}[1]{}\providecommand{\singleletter}[1]{#1}%
\begin{thebibliography}{54}%
\makeatletter
\providecommand \@ifxundefined [1]{%
 \@ifx{#1\undefined}
}%
\providecommand \@ifnum [1]{%
 \ifnum #1\expandafter \@firstoftwo
 \else \expandafter \@secondoftwo
 \fi
}%
\providecommand \@ifx [1]{%
 \ifx #1\expandafter \@firstoftwo
 \else \expandafter \@secondoftwo
 \fi
}%
\providecommand \natexlab [1]{#1}%
\providecommand \enquote  [1]{``#1''}%
\providecommand \bibnamefont  [1]{#1}%
\providecommand \bibfnamefont [1]{#1}%
\providecommand \citenamefont [1]{#1}%
\providecommand \href@noop [0]{\@secondoftwo}%
\providecommand \href [0]{\begingroup \@sanitize@url \@href}%
\providecommand \@href[1]{\@@startlink{#1}\@@href}%
\providecommand \@@href[1]{\endgroup#1\@@endlink}%
\providecommand \@sanitize@url [0]{\catcode `\\12\catcode `\$12\catcode
  `\&12\catcode `\#12\catcode `\^12\catcode `\_12\catcode `\%12\relax}%
\providecommand \@@startlink[1]{}%
\providecommand \@@endlink[0]{}%
\providecommand \url  [0]{\begingroup\@sanitize@url \@url }%
\providecommand \@url [1]{\endgroup\@href {#1}{\urlprefix }}%
\providecommand \urlprefix  [0]{URL }%
\providecommand \Eprint [0]{\href }%
\providecommand \doibase [0]{http://dx.doi.org/}%
\providecommand \selectlanguage [0]{\@gobble}%
\providecommand \bibinfo  [0]{\@secondoftwo}%
\providecommand \bibfield  [0]{\@secondoftwo}%
\providecommand \translation [1]{[#1]}%
\providecommand \BibitemOpen [0]{}%
\providecommand \bibitemStop [0]{}%
\providecommand \bibitemNoStop [0]{.\EOS\space}%
\providecommand \EOS [0]{\spacefactor3000\relax}%
\providecommand \BibitemShut  [1]{\csname bibitem#1\endcsname}%
\let\auto@bib@innerbib\@empty
\bibitem [{\citenamefont {S\'anchez}\ \emph {et~al.}(2006)\citenamefont
  {S\'anchez}, \citenamefont {N\"ortersh\"auser}, \citenamefont {Ewald} \emph
  {et~al.}}]{Li-radii2006}%
  \BibitemOpen
  \bibfield  {author} {\bibinfo {author} {\bibfnamefont {R.}~\bibnamefont
  {S\'anchez}}, \bibinfo {author} {\bibfnamefont {W.}~\bibnamefont
  {N\"ortersh\"auser}}, \bibinfo {author} {\bibfnamefont {G.}~\bibnamefont
  {Ewald}},  \emph {et~al.},\ }\href {\doibase 10.1103/PhysRevLett.96.033002}
  {\bibfield  {journal} {\bibinfo  {journal} {Phys. Rev. Lett.}\ }\textbf
  {\bibinfo {volume} {96}},\ \bibinfo {pages} {033002} (\bibinfo {year}
  {2006})}\BibitemShut {NoStop}%
\bibitem [{\citenamefont {Koszor\'us}\ \emph {et~al.}(2021)\citenamefont
  {Koszor\'us}, \citenamefont {Yang}, \citenamefont {Jiang} \emph
  {et~al.}}]{K-radii-np2021}%
  \BibitemOpen
  \bibfield  {author} {\bibinfo {author} {\bibfnamefont {A.}~\bibnamefont
  {Koszor\'us}}, \bibinfo {author} {\bibfnamefont {X.~F.}\ \bibnamefont
  {Yang}}, \bibinfo {author} {\bibfnamefont {W.~G.}\ \bibnamefont {Jiang}},
  \emph {et~al.},\ }\href {\doibase https://doi.org/10.1038/s41567-020-01136-5}
  {\bibfield  {journal} {\bibinfo  {journal} {Nat. Phys.}\ }\textbf {\bibinfo
  {volume} {17}},\ \bibinfo {pages} {439} (\bibinfo {year} {2021})}\BibitemShut
  {NoStop}%
\bibitem [{\citenamefont {Yang}\ \emph {et~al.}(2016)\citenamefont {Yang},
  \citenamefont {Wraith}, \citenamefont {Xie} \emph {et~al.}}]{79Zn-PRL2016}%
  \BibitemOpen
  \bibfield  {author} {\bibinfo {author} {\bibfnamefont {X.~F.}\ \bibnamefont
  {Yang}}, \bibinfo {author} {\bibfnamefont {C.}~\bibnamefont {Wraith}},
  \bibinfo {author} {\bibfnamefont {L.}~\bibnamefont {Xie}},  \emph {et~al.},\
  }\href {\doibase 10.1103/PhysRevLett.116.182502} {\bibfield  {journal}
  {\bibinfo  {journal} {Phys. Rev. Lett.}\ }\textbf {\bibinfo {volume} {116}},\
  \bibinfo {pages} {182502} (\bibinfo {year} {2016})}\BibitemShut {NoStop}%
\bibitem [{\citenamefont {Marsh}\ \emph {et~al.}(2018)\citenamefont {Marsh},
  \citenamefont {Goodacre}, \citenamefont {Sels} \emph
  {et~al.}}]{Hg-radii2018}%
  \BibitemOpen
  \bibfield  {author} {\bibinfo {author} {\bibfnamefont {B.~A.}\ \bibnamefont
  {Marsh}}, \bibinfo {author} {\bibfnamefont {T.~D.}\ \bibnamefont {Goodacre}},
  \bibinfo {author} {\bibfnamefont {S.}~\bibnamefont {Sels}},  \emph {et~al.},\
  }\href {\doibase 10.1038/s41567-018-0292-8} {\bibfield  {journal} {\bibinfo
  {journal} {Nat. Phys.}\ }\textbf {\bibinfo {volume} {14}},\ \bibinfo {pages}
  {1163} (\bibinfo {year} {2018})}\BibitemShut {NoStop}%
\bibitem [{\citenamefont {Ye}\ \emph {et~al.}(2025)\citenamefont {Ye},
  \citenamefont {Yang}, \citenamefont {Sakurai},\ and\ \citenamefont
  {Hu}}]{Ye2025}%
  \BibitemOpen
  \bibfield  {author} {\bibinfo {author} {\bibfnamefont {Y.}~\bibnamefont
  {Ye}}, \bibinfo {author} {\bibfnamefont {X.}~\bibnamefont {Yang}}, \bibinfo
  {author} {\bibfnamefont {H.}~\bibnamefont {Sakurai}}, \ and\ \bibinfo
  {author} {\bibfnamefont {B.}~\bibnamefont {Hu}},\ }\href {\doibase
  10.1038/s42254-024-00782-5} {\bibfield  {journal} {\bibinfo  {journal} {Nat.
  Rev. Phys.}\ }\textbf {\bibinfo {volume} {7}},\ \bibinfo {pages} {21}
  (\bibinfo {year} {2025})}\BibitemShut {NoStop}%
\bibitem [{\citenamefont {Garcia~Ruiz}\ \emph {et~al.}(2016)\citenamefont
  {Garcia~Ruiz}, \citenamefont {Bissell}, \citenamefont {Blaum} \emph
  {et~al.}}]{Ca-radii-np2016}%
  \BibitemOpen
  \bibfield  {author} {\bibinfo {author} {\bibfnamefont {R.~F.}\ \bibnamefont
  {Garcia~Ruiz}}, \bibinfo {author} {\bibfnamefont {M.~L.}\ \bibnamefont
  {Bissell}}, \bibinfo {author} {\bibfnamefont {K.}~\bibnamefont {Blaum}},
  \emph {et~al.},\ }\href {\doibase 10.1038/nphys3645} {\bibfield  {journal}
  {\bibinfo  {journal} {Nat. Phys.}\ }\textbf {\bibinfo {volume} {12}},\
  \bibinfo {pages} {594} (\bibinfo {year} {2016})}\BibitemShut {NoStop}%
\bibitem [{\citenamefont {de~Groote}\ \emph {et~al.}(2020)\citenamefont
  {de~Groote}, \citenamefont {Billowes}, \citenamefont {Binnersley} \emph
  {et~al.}}]{Cu-radii-np2020}%
  \BibitemOpen
  \bibfield  {author} {\bibinfo {author} {\bibfnamefont {R.~P.}\ \bibnamefont
  {de~Groote}}, \bibinfo {author} {\bibfnamefont {J.}~\bibnamefont {Billowes}},
  \bibinfo {author} {\bibfnamefont {C.~L.}\ \bibnamefont {Binnersley}},  \emph
  {et~al.},\ }\href {\doibase 10.1038/s41567-020-0868-y} {\bibfield  {journal}
  {\bibinfo  {journal} {Nat. Phys.}\ }\textbf {\bibinfo {volume} {16}},\
  \bibinfo {pages} {620} (\bibinfo {year} {2020})}\BibitemShut {NoStop}%
\bibitem [{\citenamefont {Malbrunot-Ettenauer}\ \emph
  {et~al.}(2022)\citenamefont {Malbrunot-Ettenauer}, \citenamefont {Kaufmann},
  \citenamefont {Bacca} \emph {et~al.}}]{Ni1-prl2022}%
  \BibitemOpen
  \bibfield  {author} {\bibinfo {author} {\bibfnamefont {S.}~\bibnamefont
  {Malbrunot-Ettenauer}}, \bibinfo {author} {\bibfnamefont {S.}~\bibnamefont
  {Kaufmann}}, \bibinfo {author} {\bibfnamefont {S.}~\bibnamefont {Bacca}},
  \emph {et~al.},\ }\href {\doibase 10.1103/PhysRevLett.128.022502} {\bibfield
  {journal} {\bibinfo  {journal} {Phys. Rev. Lett.}\ }\textbf {\bibinfo
  {volume} {128}},\ \bibinfo {pages} {022502} (\bibinfo {year}
  {2022})}\BibitemShut {NoStop}%
\bibitem [{\citenamefont {Sommer}\ \emph {et~al.}(2022)\citenamefont {Sommer},
  \citenamefont {K\"onig}, \citenamefont {Rossi} \emph {et~al.}}]{Ni2-prl2022}%
  \BibitemOpen
  \bibfield  {author} {\bibinfo {author} {\bibfnamefont {F.}~\bibnamefont
  {Sommer}}, \bibinfo {author} {\bibfnamefont {K.}~\bibnamefont {K\"onig}},
  \bibinfo {author} {\bibfnamefont {D.~M.}\ \bibnamefont {Rossi}},  \emph
  {et~al.},\ }\href {\doibase 10.1103/PhysRevLett.129.132501} {\bibfield
  {journal} {\bibinfo  {journal} {Phys. Rev. Lett.}\ }\textbf {\bibinfo
  {volume} {129}},\ \bibinfo {pages} {132501} (\bibinfo {year}
  {2022})}\BibitemShut {NoStop}%
\bibitem [{\citenamefont {Gorges}\ \emph {et~al.}(2019)\citenamefont {Gorges},
  \citenamefont {Rodr\'{\i}guez}, \citenamefont {Balabanski} \emph
  {et~al.}}]{Sn-radii-prl2019}%
  \BibitemOpen
  \bibfield  {author} {\bibinfo {author} {\bibfnamefont {C.}~\bibnamefont
  {Gorges}}, \bibinfo {author} {\bibfnamefont {L.~V.}\ \bibnamefont
  {Rodr\'{\i}guez}}, \bibinfo {author} {\bibfnamefont {D.~L.}\ \bibnamefont
  {Balabanski}},  \emph {et~al.},\ }\href {\doibase
  10.1103/PhysRevLett.122.192502} {\bibfield  {journal} {\bibinfo  {journal}
  {Phys. Rev. Lett.}\ }\textbf {\bibinfo {volume} {122}},\ \bibinfo {pages}
  {192502} (\bibinfo {year} {2019})}\BibitemShut {NoStop}%
\bibitem [{\citenamefont {Cubiss}\ \emph {et~al.}(2023)\citenamefont {Cubiss},
  \citenamefont {Andreyev}, \citenamefont {Barzakh} \emph
  {et~al.}}]{Au-radii2023}%
  \BibitemOpen
  \bibfield  {author} {\bibinfo {author} {\bibfnamefont {J.~G.}\ \bibnamefont
  {Cubiss}}, \bibinfo {author} {\bibfnamefont {A.~N.}\ \bibnamefont
  {Andreyev}}, \bibinfo {author} {\bibfnamefont {A.~E.}\ \bibnamefont
  {Barzakh}},  \emph {et~al.},\ }\href {\doibase
  10.1103/PhysRevLett.131.202501} {\bibfield  {journal} {\bibinfo  {journal}
  {Phys. Rev. Lett.}\ }\textbf {\bibinfo {volume} {131}},\ \bibinfo {pages}
  {202501} (\bibinfo {year} {2023})}\BibitemShut {NoStop}%
\bibitem [{\citenamefont {Palmer}\ \emph {et~al.}(1984)\citenamefont {Palmer},
  \citenamefont {Baird}, \citenamefont {Blundell} \emph
  {et~al.}}]{Ca-radii-1984}%
  \BibitemOpen
  \bibfield  {author} {\bibinfo {author} {\bibfnamefont {C.~W.~P.}\
  \bibnamefont {Palmer}}, \bibinfo {author} {\bibfnamefont {P.~E.~G.}\
  \bibnamefont {Baird}}, \bibinfo {author} {\bibfnamefont {S.~A.}\ \bibnamefont
  {Blundell}},  \emph {et~al.},\ }\href {\doibase 10.1088/0022-3700/17/11/014}
  {\bibfield  {journal} {\bibinfo  {journal} {J. Phys. B: Atom. Mol. Phys.}\
  }\textbf {\bibinfo {volume} {17}},\ \bibinfo {pages} {2197} (\bibinfo {year}
  {1984})}\BibitemShut {NoStop}%
\bibitem [{\citenamefont {Blaum}\ \emph {et~al.}(2008)\citenamefont {Blaum},
  \citenamefont {Geithner}, \citenamefont {Lassen} \emph
  {et~al.}}]{Ar-radii-npa2008}%
  \BibitemOpen
  \bibfield  {author} {\bibinfo {author} {\bibfnamefont {K.}~\bibnamefont
  {Blaum}}, \bibinfo {author} {\bibfnamefont {W.}~\bibnamefont {Geithner}},
  \bibinfo {author} {\bibfnamefont {J.}~\bibnamefont {Lassen}},  \emph
  {et~al.},\ }\href {\doibase https://doi.org/10.1016/j.nuclphysa.2007.11.004}
  {\bibfield  {journal} {\bibinfo  {journal} {Nucl. Phys. A}\ }\textbf
  {\bibinfo {volume} {799}},\ \bibinfo {pages} {30} (\bibinfo {year}
  {2008})}\BibitemShut {NoStop}%
\bibitem [{\citenamefont {Miller}\ \emph {et~al.}(2019)\citenamefont {Miller},
  \citenamefont {Minamisono}, \citenamefont {Klose} \emph
  {et~al.}}]{Ca-radii-np2019}%
  \BibitemOpen
  \bibfield  {author} {\bibinfo {author} {\bibfnamefont {A.~J.}\ \bibnamefont
  {Miller}}, \bibinfo {author} {\bibfnamefont {K.}~\bibnamefont {Minamisono}},
  \bibinfo {author} {\bibfnamefont {A.}~\bibnamefont {Klose}},  \emph
  {et~al.},\ }\href {\doibase 10.1038/s41567-019-0416-9} {\bibfield  {journal}
  {\bibinfo  {journal} {Nat. Phys.}\ }\textbf {\bibinfo {volume} {15}},\
  \bibinfo {pages} {432} (\bibinfo {year} {2019})}\BibitemShut {NoStop}%
\bibitem [{\citenamefont {Reinhard}\ and\ \citenamefont
  {Nazarewicz}(2021)}]{Reinhard2021}%
  \BibitemOpen
  \bibfield  {author} {\bibinfo {author} {\bibfnamefont {P.-G.}\ \bibnamefont
  {Reinhard}}\ and\ \bibinfo {author} {\bibfnamefont {W.}~\bibnamefont
  {Nazarewicz}},\ }\href {\doibase 10.1103/PhysRevC.103.054310} {\bibfield
  {journal} {\bibinfo  {journal} {Phys. Rev. C}\ }\textbf {\bibinfo {volume}
  {103}},\ \bibinfo {pages} {054310} (\bibinfo {year} {2021})}\BibitemShut
  {NoStop}%
\bibitem [{\citenamefont {Rossi}\ \emph {et~al.}(2015)\citenamefont {Rossi},
  \citenamefont {Minamisono}, \citenamefont {Asberry} \emph
  {et~al.}}]{K-radii-prc2015}%
  \BibitemOpen
  \bibfield  {author} {\bibinfo {author} {\bibfnamefont {D.~M.}\ \bibnamefont
  {Rossi}}, \bibinfo {author} {\bibfnamefont {K.}~\bibnamefont {Minamisono}},
  \bibinfo {author} {\bibfnamefont {H.~B.}\ \bibnamefont {Asberry}},  \emph
  {et~al.},\ }\href {\doibase 10.1103/PhysRevC.92.014305} {\bibfield  {journal}
  {\bibinfo  {journal} {Phys. Rev. C}\ }\textbf {\bibinfo {volume} {92}},\
  \bibinfo {pages} {014305} (\bibinfo {year} {2015})}\BibitemShut {NoStop}%
\bibitem [{\citenamefont {K\"onig}\ \emph {et~al.}(2023)\citenamefont
  {K\"onig}, \citenamefont {Fritzsche}, \citenamefont {Hagen} \emph
  {et~al.}}]{Sc-radii-prl2023}%
  \BibitemOpen
  \bibfield  {author} {\bibinfo {author} {\bibfnamefont {K.}~\bibnamefont
  {K\"onig}}, \bibinfo {author} {\bibfnamefont {S.}~\bibnamefont {Fritzsche}},
  \bibinfo {author} {\bibfnamefont {G.}~\bibnamefont {Hagen}},  \emph
  {et~al.},\ }\href {\doibase 10.1103/PhysRevLett.131.102501} {\bibfield
  {journal} {\bibinfo  {journal} {Phys. Rev. Lett.}\ }\textbf {\bibinfo
  {volume} {131}},\ \bibinfo {pages} {102501} (\bibinfo {year}
  {2023})}\BibitemShut {NoStop}%
\bibitem [{\citenamefont {Avgoulea}\ \emph {et~al.}(2011)\citenamefont
  {Avgoulea}, \citenamefont {Gangrsky}, \citenamefont {Marinova} \emph
  {et~al.}}]{Avgoulea-jpg2011}%
  \BibitemOpen
  \bibfield  {author} {\bibinfo {author} {\bibfnamefont {M.}~\bibnamefont
  {Avgoulea}}, \bibinfo {author} {\bibfnamefont {Y.~P.}\ \bibnamefont
  {Gangrsky}}, \bibinfo {author} {\bibfnamefont {K.~P.}\ \bibnamefont
  {Marinova}},  \emph {et~al.},\ }\href {\doibase
  https://doi.org/10.1088/0954-3899/38/2/025104} {\bibfield  {journal}
  {\bibinfo  {journal} {J. Phys. G: Nucl. Part. Phys.}\ }\textbf {\bibinfo
  {volume} {38}},\ \bibinfo {pages} {025104} (\bibinfo {year}
  {2011})}\BibitemShut {NoStop}%
\bibitem [{\citenamefont {$\textrm{Á. Koszorús}$}\ \emph
  {et~al.}(2021)\citenamefont {$\textrm{Á. Koszorús}$}, \citenamefont
  {Vormawah}, \citenamefont {Beerwerth} \emph {et~al.}}]{42Sc-agi}%
  \BibitemOpen
  \bibfield  {author} {\bibinfo {author} {\bibnamefont {$\textrm{Á.
  Koszorús}$}}, \bibinfo {author} {\bibfnamefont {L.}~\bibnamefont
  {Vormawah}}, \bibinfo {author} {\bibfnamefont {R.}~\bibnamefont {Beerwerth}},
   \emph {et~al.},\ }\href {\doibase
  https://doi.org/10.1016/j.physletb.2021.136439} {\bibfield  {journal}
  {\bibinfo  {journal} {Phys. Lett. B}\ }\textbf {\bibinfo {volume} {819}},\
  \bibinfo {pages} {136439} (\bibinfo {year} {2021})}\BibitemShut {NoStop}%
\bibitem [{\citenamefont {$\textrm{S. W}$. Bai}\ \emph
  {et~al.}(2022)\citenamefont {$\textrm{S. W}$. Bai}, \citenamefont
  {$\textrm{Á. Koszorús}$}, \citenamefont {$\textrm{B. S}$. Hu} \emph
  {et~al.}}]{Sc-plb}%
  \BibitemOpen
  \bibfield  {author} {\bibinfo {author} {\bibnamefont {$\textrm{S. W}$. Bai}},
  \bibinfo {author} {\bibnamefont {$\textrm{Á. Koszorús}$}}, \bibinfo
  {author} {\bibnamefont {$\textrm{B. S}$. Hu}},  \emph {et~al.},\ }\href
  {\doibase https://doi.org/10.1016/j.physletb.2022.137064} {\bibfield
  {journal} {\bibinfo  {journal} {Phys. Lett. B}\ }\textbf {\bibinfo {volume}
  {829}},\ \bibinfo {pages} {137064} (\bibinfo {year} {2022})}\BibitemShut
  {NoStop}%
\bibitem [{\citenamefont {Fedosseev}\ \emph {et~al.}(2012)\citenamefont
  {Fedosseev}, \citenamefont {Berg}, \citenamefont {Fedorov} \emph
  {et~al.}}]{RILIS}%
  \BibitemOpen
  \bibfield  {author} {\bibinfo {author} {\bibfnamefont {V.~N.}\ \bibnamefont
  {Fedosseev}}, \bibinfo {author} {\bibfnamefont {L.-E.}\ \bibnamefont {Berg}},
  \bibinfo {author} {\bibfnamefont {D.~V.}\ \bibnamefont {Fedorov}},  \emph
  {et~al.},\ }\href {\doibase 10.1063/1.3662206} {\bibfield  {journal}
  {\bibinfo  {journal} {Rev. Sci. Instrum.}\ }\textbf {\bibinfo {volume}
  {83}},\ \bibinfo {pages} {02A903} (\bibinfo {year} {2012})}\BibitemShut
  {NoStop}%
\bibitem [{\citenamefont {Mané}\ \emph {et~al.}(2009)\citenamefont {Mané},
  \citenamefont {Billowes}, \citenamefont {Campbell} \emph {et~al.}}]{ISCOOL}%
  \BibitemOpen
  \bibfield  {author} {\bibinfo {author} {\bibfnamefont {E.}~\bibnamefont
  {Mané}}, \bibinfo {author} {\bibfnamefont {J.}~\bibnamefont {Billowes}},
  \bibinfo {author} {\bibfnamefont {P.}~\bibnamefont {Campbell}},  \emph
  {et~al.},\ }\href {\doibase 10.1140/epja/i2009-10828-0} {\bibfield  {journal}
  {\bibinfo  {journal} {Eur. Phys. J. A}\ }\textbf {\bibinfo {volume} {42}},\
  \bibinfo {pages} {503} (\bibinfo {year} {2009})}\BibitemShut {NoStop}%
\bibitem [{\citenamefont {Visscher}\ \emph {et~al.}(2001)\citenamefont
  {Visscher}, \citenamefont {Eliav},\ and\ \citenamefont {Kaldor}}]{FSCC-2001}%
  \BibitemOpen
  \bibfield  {author} {\bibinfo {author} {\bibfnamefont {L.}~\bibnamefont
  {Visscher}}, \bibinfo {author} {\bibfnamefont {E.}~\bibnamefont {Eliav}}, \
  and\ \bibinfo {author} {\bibfnamefont {U.}~\bibnamefont {Kaldor}},\ }\href
  {\doibase 10.1063/1.1415746} {\bibfield  {journal} {\bibinfo  {journal} {J.
  Chem. Phys.}\ }\textbf {\bibinfo {volume} {115}},\ \bibinfo {pages} {9720}
  (\bibinfo {year} {2001})}\BibitemShut {NoStop}%
\bibitem [{\citenamefont {Dzuba}\ \emph {et~al.}(1996)\citenamefont {Dzuba},
  \citenamefont {Flambaum},\ and\ \citenamefont {Kozlov}}]{CIMBPT}%
  \BibitemOpen
  \bibfield  {author} {\bibinfo {author} {\bibfnamefont {V.~A.}\ \bibnamefont
  {Dzuba}}, \bibinfo {author} {\bibfnamefont {V.~V.}\ \bibnamefont {Flambaum}},
  \ and\ \bibinfo {author} {\bibfnamefont {M.~G.}\ \bibnamefont {Kozlov}},\
  }\href {\doibase 10.1103/PhysRevA.54.3948} {\bibfield  {journal} {\bibinfo
  {journal} {Phys. Rev. A}\ }\textbf {\bibinfo {volume} {54}},\ \bibinfo
  {pages} {3948} (\bibinfo {year} {1996})}\BibitemShut {NoStop}%
\bibitem [{\citenamefont {Gins}\ \emph {et~al.}(2018)\citenamefont {Gins},
  \citenamefont {{de Groote}}, \citenamefont {Bissell} \emph
  {et~al.}}]{SATLAS}%
  \BibitemOpen
  \bibfield  {author} {\bibinfo {author} {\bibfnamefont {W.}~\bibnamefont
  {Gins}}, \bibinfo {author} {\bibfnamefont {R.}~\bibnamefont {{de Groote}}},
  \bibinfo {author} {\bibfnamefont {M.}~\bibnamefont {Bissell}},  \emph
  {et~al.},\ }\href {\doibase https://doi.org/10.1016/j.cpc.2017.09.012}
  {\bibfield  {journal} {\bibinfo  {journal} {Comput. Phys. Commun.}\ }\textbf
  {\bibinfo {volume} {222}},\ \bibinfo {pages} {286 } (\bibinfo {year}
  {2018})}\BibitemShut {NoStop}%
\bibitem [{\citenamefont {Kaldor}\ and\ \citenamefont
  {Eliav}(1998)}]{FSCC-1998}%
  \BibitemOpen
  \bibfield  {author} {\bibinfo {author} {\bibfnamefont {U.}~\bibnamefont
  {Kaldor}}\ and\ \bibinfo {author} {\bibfnamefont {E.}~\bibnamefont {Eliav}},\
  }\href {\doibase https://doi.org/10.1016/S0065-3276(08)60194-X} {\ \bibinfo
  {series} {Adv. Quantum Chem.},\ \textbf {\bibinfo {volume} {31}},\ \bibinfo
  {pages} {313} (\bibinfo {year} {1998})}\BibitemShut {NoStop}%
\bibitem [{\citenamefont {Gomes}\ \emph {et~al.}(2019)\citenamefont {Gomes},
  \citenamefont {Saue}, \citenamefont {Visscher} \emph {et~al.}}]{DIRAC19}%
  \BibitemOpen
  \bibfield  {author} {\bibinfo {author} {\bibfnamefont {A.~S.~P.}\
  \bibnamefont {Gomes}}, \bibinfo {author} {\bibfnamefont {T.}~\bibnamefont
  {Saue}}, \bibinfo {author} {\bibfnamefont {L.}~\bibnamefont {Visscher}},
  \emph {et~al.},\ }\href {\doibase 10.5281/zenodo.3572669} {\enquote {\bibinfo
  {title} {$\textrm{DIRAC19}$},}\ } (\bibinfo {year} {2019})\BibitemShut
  {NoStop}%
\bibitem [{\citenamefont {Kahl}\ and\ \citenamefont
  {Berengut}(2019)}]{AMBIT-2019}%
  \BibitemOpen
  \bibfield  {author} {\bibinfo {author} {\bibfnamefont {E.}~\bibnamefont
  {Kahl}}\ and\ \bibinfo {author} {\bibfnamefont {J.}~\bibnamefont
  {Berengut}},\ }\href {\doibase https://doi.org/10.1016/j.cpc.2018.12.014}
  {\bibfield  {journal} {\bibinfo  {journal} {Comput. Phys. Commun.}\ }\textbf
  {\bibinfo {volume} {238}},\ \bibinfo {pages} {232} (\bibinfo {year}
  {2019})}\BibitemShut {NoStop}%
\bibitem [{\citenamefont {Landau}\ \emph {et~al.}(2001)\citenamefont {Landau},
  \citenamefont {Eliav},\ and\ \citenamefont
  {Kaldor}}]{landau2001intermediate}%
  \BibitemOpen
  \bibfield  {author} {\bibinfo {author} {\bibfnamefont {A.}~\bibnamefont
  {Landau}}, \bibinfo {author} {\bibfnamefont {E.}~\bibnamefont {Eliav}}, \
  and\ \bibinfo {author} {\bibfnamefont {U.}~\bibnamefont {Kaldor}},\
  }\href@noop {} {\bibfield  {journal} {\bibinfo  {journal} {Adv. Quantum
  Chem.}\ }\textbf {\bibinfo {volume} {39}},\ \bibinfo {pages} {171} (\bibinfo
  {year} {2001})}\BibitemShut {NoStop}%
\bibitem [{\citenamefont {Dyall}\ and\ \citenamefont {Gomes}()}]{dyallprep3d}%
  \BibitemOpen
  \bibfield  {author} {\bibinfo {author} {\bibfnamefont {K.}~\bibnamefont
  {Dyall}}\ and\ \bibinfo {author} {\bibfnamefont {A.}~\bibnamefont {Gomes}},\
  }\href@noop {} {\bibinfo  {journal} {manuscript in preparation}\
  }\BibitemShut {NoStop}%
\bibitem [{\citenamefont {Visscher}\ and\ \citenamefont
  {Dyall}(1997)}]{visscher-nuccharge}%
  \BibitemOpen
\bibfield  {journal} {  }\bibfield  {author} {\bibinfo {author} {\bibfnamefont
  {L.}~\bibnamefont {Visscher}}\ and\ \bibinfo {author} {\bibfnamefont
  {K.}~\bibnamefont {Dyall}},\ }\href {\doibase
  https://doi.org/10.1006/adnd.1997.0751} {\bibfield  {journal} {\bibinfo
  {journal} {At. Data Nucl. Data Tables}\ }\textbf {\bibinfo {volume} {67}},\
  \bibinfo {pages} {207} (\bibinfo {year} {1997})}\BibitemShut {NoStop}%
\bibitem [{\citenamefont {Gustafsson}\ \emph {et~al.}(2020)\citenamefont
  {Gustafsson}, \citenamefont {Ricketts}, \citenamefont {Reitsma} \emph
  {et~al.}}]{Sn-Fredik}%
  \BibitemOpen
  \bibfield  {author} {\bibinfo {author} {\bibfnamefont {F.~P.}\ \bibnamefont
  {Gustafsson}}, \bibinfo {author} {\bibfnamefont {C.~M.}\ \bibnamefont
  {Ricketts}}, \bibinfo {author} {\bibfnamefont {M.~L.}\ \bibnamefont
  {Reitsma}},  \emph {et~al.},\ }\href {\doibase 10.1103/PhysRevA.102.052812}
  {\bibfield  {journal} {\bibinfo  {journal} {Phys. Rev. A}\ }\textbf {\bibinfo
  {volume} {102}},\ \bibinfo {pages} {052812} (\bibinfo {year}
  {2020})}\BibitemShut {NoStop}%
\bibitem [{\citenamefont {Kanellakopoulos}\ \emph {et~al.}(2020)\citenamefont
  {Kanellakopoulos}, \citenamefont {Yang}, \citenamefont {Bissell} \emph
  {et~al.}}]{Ge-moment-prc2020}%
  \BibitemOpen
  \bibfield  {author} {\bibinfo {author} {\bibfnamefont {A.}~\bibnamefont
  {Kanellakopoulos}}, \bibinfo {author} {\bibfnamefont {X.~F.}\ \bibnamefont
  {Yang}}, \bibinfo {author} {\bibfnamefont {M.~L.}\ \bibnamefont {Bissell}},
  \emph {et~al.},\ }\href {\doibase 10.1103/PhysRevC.102.054331} {\bibfield
  {journal} {\bibinfo  {journal} {Phys. Rev. C}\ }\textbf {\bibinfo {volume}
  {102}},\ \bibinfo {pages} {054331} (\bibinfo {year} {2020})}\BibitemShut
  {NoStop}%
\bibitem [{\citenamefont {Udrescu}\ \emph {et~al.}(2021)\citenamefont
  {Udrescu}, \citenamefont {Brinson}, \citenamefont {Garcia~Ruiz} \emph
  {et~al.}}]{RaF-prl}%
  \BibitemOpen
  \bibfield  {author} {\bibinfo {author} {\bibfnamefont {S.~M.}\ \bibnamefont
  {Udrescu}}, \bibinfo {author} {\bibfnamefont {A.~J.}\ \bibnamefont
  {Brinson}}, \bibinfo {author} {\bibfnamefont {R.~F.}\ \bibnamefont
  {Garcia~Ruiz}},  \emph {et~al.},\ }\href {\doibase
  10.1103/PhysRevLett.127.033001} {\bibfield  {journal} {\bibinfo  {journal}
  {Phys. Rev. Lett.}\ }\textbf {\bibinfo {volume} {127}},\ \bibinfo {pages}
  {033001} (\bibinfo {year} {2021})}\BibitemShut {NoStop}%
\bibitem [{\citenamefont {Angeli}\ and\ \citenamefont
  {Marinova}(2013)}]{Radii2013}%
  \BibitemOpen
  \bibfield  {author} {\bibinfo {author} {\bibfnamefont {I.}~\bibnamefont
  {Angeli}}\ and\ \bibinfo {author} {\bibfnamefont {K.}~\bibnamefont
  {Marinova}},\ }\href {\doibase https://doi.org/10.1016/j.adt.2011.12.006}
  {\bibfield  {journal} {\bibinfo  {journal} {At. Data Nucl. Data Tables}\
  }\textbf {\bibinfo {volume} {99}},\ \bibinfo {pages} {69} (\bibinfo {year}
  {2013})}\BibitemShut {NoStop}%
\bibitem [{\citenamefont {Gangrsky}\ \emph {et~al.}(2006)\citenamefont
  {Gangrsky}, \citenamefont {Marinova}, \citenamefont {Zemlyanoi} \emph
  {et~al.}}]{IGISOL-Sc}%
  \BibitemOpen
  \bibfield  {author} {\bibinfo {author} {\bibfnamefont {Y.}~\bibnamefont
  {Gangrsky}}, \bibinfo {author} {\bibfnamefont {K.}~\bibnamefont {Marinova}},
  \bibinfo {author} {\bibfnamefont {S.}~\bibnamefont {Zemlyanoi}},  \emph
  {et~al.},\ }\href {\doibase 10.1007/s10751-006-9488-x} {\bibfield  {journal}
  {\bibinfo  {journal} {Hyperfine Interact.}\ }\textbf {\bibinfo {volume}
  {171}},\ \bibinfo {pages} {209} (\bibinfo {year} {2006})}\BibitemShut
  {NoStop}%
\bibitem [{\citenamefont {Reinhard}\ and\ \citenamefont
  {Nazarewicz}(2017)}]{Fayan-Ca-prc2017}%
  \BibitemOpen
  \bibfield  {author} {\bibinfo {author} {\bibfnamefont {P.-G.}\ \bibnamefont
  {Reinhard}}\ and\ \bibinfo {author} {\bibfnamefont {W.}~\bibnamefont
  {Nazarewicz}},\ }\href {\doibase 10.1103/PhysRevC.95.064328} {\bibfield
  {journal} {\bibinfo  {journal} {Phys. Rev. C}\ }\textbf {\bibinfo {volume}
  {95}},\ \bibinfo {pages} {064328} (\bibinfo {year} {2017})}\BibitemShut
  {NoStop}%
\bibitem [{\citenamefont {Karthein}\ \emph {et~al.}(2024)\citenamefont
  {Karthein}, \citenamefont {Ricketts}, \citenamefont {Ruiz} \emph
  {et~al.}}]{Fy-IVP-2024}%
  \BibitemOpen
  \bibfield  {author} {\bibinfo {author} {\bibfnamefont {J.}~\bibnamefont
  {Karthein}}, \bibinfo {author} {\bibfnamefont {C.~M.}\ \bibnamefont
  {Ricketts}}, \bibinfo {author} {\bibfnamefont {R.~F.~G.}\ \bibnamefont
  {Ruiz}},  \emph {et~al.},\ }\href {\doibase 10.1038/s41567-024-02612-y}
  {\bibfield  {journal} {\bibinfo  {journal} {Nat. Phys.}\ }\textbf {\bibinfo
  {volume} {20}},\ \bibinfo {pages} {1719} (\bibinfo {year}
  {2024})}\BibitemShut {NoStop}%
\bibitem [{\citenamefont {Hammen}\ \emph {et~al.}(2018)\citenamefont {Hammen},
  \citenamefont {N\"ortersh\"auser}, \citenamefont {Balabanski} \emph
  {et~al.}}]{Cd-prl2018}%
  \BibitemOpen
  \bibfield  {author} {\bibinfo {author} {\bibfnamefont {M.}~\bibnamefont
  {Hammen}}, \bibinfo {author} {\bibfnamefont {W.}~\bibnamefont
  {N\"ortersh\"auser}}, \bibinfo {author} {\bibfnamefont {D.~L.}\ \bibnamefont
  {Balabanski}},  \emph {et~al.},\ }\href {\doibase
  10.1103/PhysRevLett.121.102501} {\bibfield  {journal} {\bibinfo  {journal}
  {Phys. Rev. Lett.}\ }\textbf {\bibinfo {volume} {121}},\ \bibinfo {pages}
  {102501} (\bibinfo {year} {2018})}\BibitemShut {NoStop}%
\bibitem [{\citenamefont {Wang}\ \emph {et~al.}(2024)\citenamefont {Wang},
  \citenamefont {Kanellakopoulos}, \citenamefont {Yang} \emph
  {et~al.}}]{Ge-radii-2024}%
  \BibitemOpen
  \bibfield  {author} {\bibinfo {author} {\bibfnamefont {S.}~\bibnamefont
  {Wang}}, \bibinfo {author} {\bibfnamefont {A.}~\bibnamefont
  {Kanellakopoulos}}, \bibinfo {author} {\bibfnamefont {X.}~\bibnamefont
  {Yang}},  \emph {et~al.},\ }\href {\doibase
  https://doi.org/10.1016/j.physletb.2024.138867} {\bibfield  {journal}
  {\bibinfo  {journal} {Phys. Lett. B}\ }\textbf {\bibinfo {volume} {856}},\
  \bibinfo {pages} {138867} (\bibinfo {year} {2024})}\BibitemShut {NoStop}%
\bibitem [{\citenamefont {Geldhof}\ \emph {et~al.}(2022)\citenamefont
  {Geldhof}, \citenamefont {Kortelainen}, \citenamefont {Beliuskina} \emph
  {et~al.}}]{Pd-prl2022}%
  \BibitemOpen
  \bibfield  {author} {\bibinfo {author} {\bibfnamefont {S.}~\bibnamefont
  {Geldhof}}, \bibinfo {author} {\bibfnamefont {M.}~\bibnamefont
  {Kortelainen}}, \bibinfo {author} {\bibfnamefont {O.}~\bibnamefont
  {Beliuskina}},  \emph {et~al.},\ }\href {\doibase
  10.1103/PhysRevLett.128.152501} {\bibfield  {journal} {\bibinfo  {journal}
  {Phys. Rev. Lett.}\ }\textbf {\bibinfo {volume} {128}},\ \bibinfo {pages}
  {152501} (\bibinfo {year} {2022})}\BibitemShut {NoStop}%
\bibitem [{\citenamefont {Jiang}\ \emph {et~al.}(2020)\citenamefont {Jiang},
  \citenamefont {Ekstr\"om}, \citenamefont {Forss\'en} \emph
  {et~al.}}]{NNLOgo}%
  \BibitemOpen
  \bibfield  {author} {\bibinfo {author} {\bibfnamefont {W.~G.}\ \bibnamefont
  {Jiang}}, \bibinfo {author} {\bibfnamefont {A.}~\bibnamefont {Ekstr\"om}},
  \bibinfo {author} {\bibfnamefont {C.}~\bibnamefont {Forss\'en}},  \emph
  {et~al.},\ }\href {\doibase 10.1103/PhysRevC.102.054301} {\bibfield
  {journal} {\bibinfo  {journal} {Phys. Rev. C}\ }\textbf {\bibinfo {volume}
  {102}},\ \bibinfo {pages} {054301} (\bibinfo {year} {2020})}\BibitemShut
  {NoStop}%
\bibitem [{\citenamefont {Buskirk}\ \emph {et~al.}(2024)\citenamefont
  {Buskirk}, \citenamefont {Godbey}, \citenamefont {Nazarewicz},\ and\
  \citenamefont {Satu\l{}a}}]{energy-gap}%
  \BibitemOpen
  \bibfield  {author} {\bibinfo {author} {\bibfnamefont {L.}~\bibnamefont
  {Buskirk}}, \bibinfo {author} {\bibfnamefont {K.}~\bibnamefont {Godbey}},
  \bibinfo {author} {\bibfnamefont {W.}~\bibnamefont {Nazarewicz}}, \ and\
  \bibinfo {author} {\bibfnamefont {W.}~\bibnamefont {Satu\l{}a}},\ }\href
  {\doibase 10.1103/PhysRevC.109.044311} {\bibfield  {journal} {\bibinfo
  {journal} {Phys. Rev. C}\ }\textbf {\bibinfo {volume} {109}},\ \bibinfo
  {pages} {044311} (\bibinfo {year} {2024})}\BibitemShut {NoStop}%
\bibitem [{\citenamefont {Dobaczewski}\ \emph {et~al.}(2001)\citenamefont
  {Dobaczewski}, \citenamefont {Magierski}, \citenamefont {Nazarewicz},
  \citenamefont {Satu\l{}a},\ and\ \citenamefont {Szyma\ifmmode~\acute{n}\else
  \'{n}\fi{}ski}}]{Dobaczewski2001}%
  \BibitemOpen
  \bibfield  {author} {\bibinfo {author} {\bibfnamefont {J.}~\bibnamefont
  {Dobaczewski}}, \bibinfo {author} {\bibfnamefont {P.}~\bibnamefont
  {Magierski}}, \bibinfo {author} {\bibfnamefont {W.}~\bibnamefont
  {Nazarewicz}}, \bibinfo {author} {\bibfnamefont {W.}~\bibnamefont
  {Satu\l{}a}}, \ and\ \bibinfo {author} {\bibfnamefont {Z.}~\bibnamefont
  {Szyma\ifmmode~\acute{n}\else \'{n}\fi{}ski}},\ }\href {\doibase
  10.1103/PhysRevC.63.024308} {\bibfield  {journal} {\bibinfo  {journal} {Phys.
  Rev. C}\ }\textbf {\bibinfo {volume} {63}},\ \bibinfo {pages} {024308}
  (\bibinfo {year} {2001})}\BibitemShut {NoStop}%
\bibitem [{\citenamefont {Wang}\ \emph {et~al.}(2021)\citenamefont {Wang},
  \citenamefont {Huang}, \citenamefont {Kondev}, \citenamefont {Audi},\ and\
  \citenamefont {Naimi}}]{AME2020}%
  \BibitemOpen
  \bibfield  {author} {\bibinfo {author} {\bibfnamefont {M.}~\bibnamefont
  {Wang}}, \bibinfo {author} {\bibfnamefont {W.}~\bibnamefont {Huang}},
  \bibinfo {author} {\bibfnamefont {F.}~\bibnamefont {Kondev}}, \bibinfo
  {author} {\bibfnamefont {G.}~\bibnamefont {Audi}}, \ and\ \bibinfo {author}
  {\bibfnamefont {S.}~\bibnamefont {Naimi}},\ }\href {\doibase
  10.1088/1674-1137/abddaf} {\bibfield  {journal} {\bibinfo  {journal} {Chin.
  Phys. C}\ }\textbf {\bibinfo {volume} {45}},\ \bibinfo {pages} {030003}
  (\bibinfo {year} {2021})}\BibitemShut {NoStop}%
\bibitem [{\citenamefont {Garcia~Ruiz}\ \emph {et~al.}(2015)\citenamefont
  {Garcia~Ruiz}, \citenamefont {Bissell}, \citenamefont {Blaum} \emph
  {et~al.}}]{Ca-moments2015}%
  \BibitemOpen
  \bibfield  {author} {\bibinfo {author} {\bibfnamefont {R.~F.}\ \bibnamefont
  {Garcia~Ruiz}}, \bibinfo {author} {\bibfnamefont {M.~L.}\ \bibnamefont
  {Bissell}}, \bibinfo {author} {\bibfnamefont {K.}~\bibnamefont {Blaum}},
  \emph {et~al.},\ }\href {\doibase 10.1103/PhysRevC.91.041304} {\bibfield
  {journal} {\bibinfo  {journal} {Phys. Rev. C}\ }\textbf {\bibinfo {volume}
  {91}},\ \bibinfo {pages} {041304(R)} (\bibinfo {year} {2015})}\BibitemShut
  {NoStop}%
\bibitem [{\citenamefont {Yang}\ \emph {et~al.}(2023)\citenamefont {Yang},
  \citenamefont {Wang}, \citenamefont {Wilkins} \emph
  {et~al.}}]{Yang-ppnp2023}%
  \BibitemOpen
  \bibfield  {author} {\bibinfo {author} {\bibfnamefont {X.~F.}\ \bibnamefont
  {Yang}}, \bibinfo {author} {\bibfnamefont {S.~J.}\ \bibnamefont {Wang}},
  \bibinfo {author} {\bibfnamefont {S.~G.}\ \bibnamefont {Wilkins}},  \emph
  {et~al.},\ }\href {\doibase https://doi.org/10.1016/j.ppnp.2022.104005}
  {\bibfield  {journal} {\bibinfo  {journal} {Prog. Part. Nucl. Phys.}\
  }\textbf {\bibinfo {volume} {129}},\ \bibinfo {pages} {104005} (\bibinfo
  {year} {2023})}\BibitemShut {NoStop}%
\bibitem [{\citenamefont {Heylen}\ \emph {et~al.}(2016)\citenamefont {Heylen},
  \citenamefont {Babcock}, \citenamefont {Beerwerth} \emph
  {et~al.}}]{Mn-radii}%
  \BibitemOpen
  \bibfield  {author} {\bibinfo {author} {\bibfnamefont {H.}~\bibnamefont
  {Heylen}}, \bibinfo {author} {\bibfnamefont {C.}~\bibnamefont {Babcock}},
  \bibinfo {author} {\bibfnamefont {R.}~\bibnamefont {Beerwerth}},  \emph
  {et~al.},\ }\href {\doibase 10.1103/PhysRevC.94.054321} {\bibfield  {journal}
  {\bibinfo  {journal} {Phys. Rev. C}\ }\textbf {\bibinfo {volume} {94}},\
  \bibinfo {pages} {054321} (\bibinfo {year} {2016})}\BibitemShut {NoStop}%
\bibitem [{\citenamefont {Zamick}(1971)}]{Zamick1971}%
  \BibitemOpen
  \bibfield  {author} {\bibinfo {author} {\bibfnamefont {L.}~\bibnamefont
  {Zamick}},\ }\href {\doibase 10.1016/0003-4916(71)90080-7} {\bibfield
  {journal} {\bibinfo  {journal} {Ann. Phys.}\ }\textbf {\bibinfo {volume}
  {66}},\ \bibinfo {pages} {784 } (\bibinfo {year} {1971})}\BibitemShut
  {NoStop}%
\bibitem [{\citenamefont {Talmi}(1984)}]{Talmi1984}%
  \BibitemOpen
  \bibfield  {author} {\bibinfo {author} {\bibfnamefont {I.}~\bibnamefont
  {Talmi}},\ }\href {\doibase 10.1016/0375-9474(84)90587-6} {\bibfield
  {journal} {\bibinfo  {journal} {Nucl. Phys. A}\ }\textbf {\bibinfo {volume}
  {423}},\ \bibinfo {pages} {189 } (\bibinfo {year} {1984})}\BibitemShut
  {NoStop}%
\bibitem [{\citenamefont {Kortelainen}\ \emph {et~al.}(2022)\citenamefont
  {Kortelainen}, \citenamefont {Sun}, \citenamefont {Hagen} \emph
  {et~al.}}]{Ca-Zn-prc2022}%
  \BibitemOpen
  \bibfield  {author} {\bibinfo {author} {\bibfnamefont {M.}~\bibnamefont
  {Kortelainen}}, \bibinfo {author} {\bibfnamefont {Z.}~\bibnamefont {Sun}},
  \bibinfo {author} {\bibfnamefont {G.}~\bibnamefont {Hagen}},  \emph
  {et~al.},\ }\href {\doibase 10.1103/PhysRevC.105.L021303} {\bibfield
  {journal} {\bibinfo  {journal} {Phys. Rev. C}\ }\textbf {\bibinfo {volume}
  {105}},\ \bibinfo {pages} {L021303} (\bibinfo {year} {2022})}\BibitemShut
  {NoStop}%
\bibitem [{\citenamefont {Brown}\ and\ \citenamefont
  {Minamisono}(2022)}]{beta-corr-prc2022}%
  \BibitemOpen
  \bibfield  {author} {\bibinfo {author} {\bibfnamefont {B.~A.}\ \bibnamefont
  {Brown}}\ and\ \bibinfo {author} {\bibfnamefont {K.}~\bibnamefont
  {Minamisono}},\ }\href {\doibase 10.1103/PhysRevC.106.L011304} {\bibfield
  {journal} {\bibinfo  {journal} {Phys. Rev. C}\ }\textbf {\bibinfo {volume}
  {106}},\ \bibinfo {pages} {L011304} (\bibinfo {year} {2022})}\BibitemShut
  {NoStop}%
\bibitem [{\citenamefont {Emrich}\ \emph {et~al.}(1983)\citenamefont {Emrich},
  \citenamefont {Fricke}, \citenamefont {Mallot}, \citenamefont {Miska},
  \citenamefont {Sieberling}, \citenamefont {Cavedon}, \citenamefont {Frois},\
  and\ \citenamefont {Goutte}}]{Emrich1983}%
  \BibitemOpen
  \bibfield  {author} {\bibinfo {author} {\bibfnamefont {H.}~\bibnamefont
  {Emrich}}, \bibinfo {author} {\bibfnamefont {G.}~\bibnamefont {Fricke}},
  \bibinfo {author} {\bibfnamefont {G.}~\bibnamefont {Mallot}}, \bibinfo
  {author} {\bibfnamefont {H.}~\bibnamefont {Miska}}, \bibinfo {author}
  {\bibfnamefont {H.-G.}\ \bibnamefont {Sieberling}}, \bibinfo {author}
  {\bibfnamefont {J.}~\bibnamefont {Cavedon}}, \bibinfo {author} {\bibfnamefont
  {B.}~\bibnamefont {Frois}}, \ and\ \bibinfo {author} {\bibfnamefont
  {D.}~\bibnamefont {Goutte}},\ }\href {\doibase 10.1016/0375-9474(83)90034-9}
  {\bibfield  {journal} {\bibinfo  {journal} {Nucl. Phys. A}\ }\textbf
  {\bibinfo {volume} {396}},\ \bibinfo {pages} {401 } (\bibinfo {year}
  {1983})}\BibitemShut {NoStop}%
\bibitem [{\citenamefont {Hu}\ \emph {et~al.}(2022)\citenamefont {Hu},
  \citenamefont {Jiang}, \citenamefont {Miyagi} \emph {et~al.}}]{Hu2022}%
  \BibitemOpen
  \bibfield  {author} {\bibinfo {author} {\bibfnamefont {B.~S.}\ \bibnamefont
  {Hu}}, \bibinfo {author} {\bibfnamefont {W.~G.}\ \bibnamefont {Jiang}},
  \bibinfo {author} {\bibfnamefont {T.}~\bibnamefont {Miyagi}},  \emph
  {et~al.},\ }\href {\doibase 10.1038/s41567-023-02324-9} {\bibfield  {journal}
  {\bibinfo  {journal} {Nat. Phys.}\ }\textbf {\bibinfo {volume} {18}},\
  \bibinfo {pages} {1196} (\bibinfo {year} {2022})}\BibitemShut {NoStop}%
\end{thebibliography}%

\end{document}